\documentclass[aps,pre,longbibliography,notitlepage,floatfix,nofootinbib]{revtex4-1}

\usepackage{bm,amsmath,amssymb,graphicx,stmaryrd,float,subcaption,upgreek,sidecap,dsfont,mathtools,gensymb}
\usepackage[abs]{overpic}
\newcommand{\uvc}[1]{\bm{\mathrm{\hat #1}}} 

\usepackage{comment,footnote}
\usepackage[format=plain,labelfont={bf,small},textfont=small,justification=raggedright,singlelinecheck=false]{caption}

\usepackage{xcolor}

\newcommand{\bdelta}{{\tilde{\delta}}}
\newcommand{\JI}{{J^{-1}}}
\newcommand{\JSI}{{J_S^{-1}}}
\newcommand{\Lgeom}{{\mathcal{L}}}

\newcommand{\dt}{{d_t}}
\newcommand{\ds}{{d_s}}
\newcommand{\rconfig}[1]{{\bar{#1}}}
\newcommand{\mc}{{\eta}}
\newcommand{\Lref}{{\bar{\mathcal{L}}}}
\newcommand{\g}{{g}}
\newcommand{\ga}{{a}}

\newcommand{\bx}{{\bf x}}

\newcommand{\shift}[1]{#1'}
\newcommand{\shop}[1]{\tilde{#1}}
\newcommand{\dvol}{dV}
\newcommand{\dvolref}{d\bar{V}}
\newcommand{\dArea}{dA}

\newcommand{\BH}{{\scriptstyle B}} 

\newcommand{\Grad}{{\rconfig{\nabla}}}
\newcommand{\grad}{{\nabla}}
\newcommand{\F}{{\bf F}}

\newcommand{\bomega}{\mbox{\boldmath $\omega$}}
\newcommand{\bOmega}{\mbox{\boldmath $\Omega$}}
\newcommand{\bn}{{\bm n}}
\newcommand{\bmo}{{\bm m}}

\newcommand{\bD}{{\bf D}}
\newcommand{\bF}{{\bf F}}
\newcommand{\bM}{{\bf M}}
\newcommand{\bff}{{\bf f}}
\newcommand{\bfm}{{\bf m}}
\newcommand{\bv}{{\bm v}}

\newcommand{\bR}{{\bf R}}
\newcommand{\bJ}{{\bf J}}

\newcommand{\jump}[1]{\left\llbracket#1\right\rrbracket}

\newcommand{\EQ}{\mathcal{E}_{(\mathcal{Q})}}
\newcommand{\EJ}{\mathcal{E}_{(\mathcal{J})}}
\newcommand{\vEJ}{\bm{\mathcal{E}}_{(\mathcal{J})}}

\begin{document}
	\title{Pseudomomentum: origins and consequences}
	\author{H. Singh\footnote{Corresponding author}}
	\email{harmeet.singh@epfl.ch}
	\affiliation{Institute of Mathematics, Laboratory for Computation and Visualisation in Mathematics and Mechanics, \'Ecole polytechnique f\'ed\'erale de Lausanne,  MA C1 612 (B\^atiment MA), Station 8, CH-1015 Lausanne, Switzerland.}
	\author{J. A. Hanna}
	\email{jhanna@unr.edu}
	\affiliation{Department of Mechanical Engineering, University of Nevada, 1664 N. Virginia St.\ (0312), Reno, NV 89557-0312, U.S.A.}
	\date{\today}

\begin{abstract}
The balance of pseudomomentum is discussed and applied to simple elasticity, ideal fluids, and the mechanics of inextensible rods and sheets.
A general framework is presented in which the simultaneous variation of an action with respect to position, time, and material labels yields bulk balance laws and jump conditions for momentum, energy, and pseudomomentum. 
The example of simple elasticity of space-filling solids is treated at length.
The pseudomomentum balance in ideal fluids is shown to imply conservation of vorticity, circulation, and helicity, and a mathematical similarity is noted between the evaluation of circulation along a material loop and the J-integral of fracture mechanics.
Integration of the pseudomomentum balance, making use of a prescription for singular sources derived by analogy with the continuous form of the balance, directly provides the propulsive force driving passive reconfiguration or locomotion of confined, inhomogeneous elastic rods.
The conserved angular momentum and pseudomomentum are identified in the classification of conical sheets with rotational inertia or bending energy.
\end{abstract}
\maketitle

\vspace{-0.5cm}
\section{Introduction}\label{sec:introduction}

Pseudomomentum is a property of material continua.  It is variously referred to as material momentum, configurational momentum, Eshelbian momentum, quasimomentum, and impulse.  Our impression is that its balance law is under-utilized and under-recognized in many branches of continuum and structural mechanics.  The current effort is our attempt to synthesize and expand a body of prior work on variational approaches and symmetry in continua, as well as revisit and reinterpret several results in the mechanics of thin structures.

While a material element acquires momentum by virtue of its motion in space, pseudomomentum is a property associated with the motion of an element or defect through the material medium itself. 
As Peierls \cite{peierls1985} commented, the distinction between the momentum and the pseudomomentum of an electromagnetic or other field is only meaningful in the presence of a material medium.
An important early paper in this area is that of Rogula \cite{rogula1966}, who defined a Lagrangian density of fields interacting with a homogeneous body, and constructed unnamed conservation laws associated with shifts in material labels.  He remarked on the difference between fields defined per unit mass of material and those associated with the embedding space.  
The influential work of Eshelby defined the ``force" on an elastic singularity as a dynamic quantity conjugate to a kinematic parameter characterizing the position of the defect within the material medium \cite{eshelby51}. This material or configurational ``force" and the pseudomomentum have the same relationship to each other as the standard force and momentum.\
 Later, Eshelby introduced a four-dimensional elastic ``energy-momentum tensor" \cite{eshelby1970,eshelby75} by analogy with the energy-momentum tensors of classical field theories \cite{landaulifshitzfields1971}.  The three-dimensional non-temporal part of this tensor has come to be known as the Eshelby tensor; its elements are material components corresponding to pseudomomentum rather than spatial components corresponding to momentum as in the classical construction\footnote{Whereas some fields such as gravity are associated with space-time, other fields such as displacements in elasticity are attached to a set of material elements. The state of a body is described using both material coordinates describing locations within the medium and spatial coordinates associated with the embedding of the body in the ambient space.  The non-temporal part of an ``energy-momentum tensor'' can be obtained when the action is varied with respect to either material or spatial coordinates, yielding either the Eshelbian or classical result.  These are different physical quantities.  As the flux associated with momentum is known as ``stress'' in continuum mechanics, that associated with pseudomomentum (material momentum) could be called pseudostress (material stress).  Although, as Ericksen \cite{Ericksen95} points out, Eshelby's exposition was in a linear-elastic context, where the distinction between spatial and material coordinates is blurred.}. 
The Eshelby tensor is related to the path independent integrals of Rice and Cherepanov in fracture mechanics \cite{rice1968, Cherepanov1967}.
Rogula later revisited material forces and corresponding balance laws in the presence of localized inhomogeneities \cite{rogula77}.
In a series of papers, A. Golebiewska Herrmann presented a unified Lagrangian treatment of continuum mechanics for systems that admit an action principle, 
deriving a complete set of dynamic balance laws
\cite{herrmannalicia1981,herrmannalicia1982,herrmannalicia1983}. 
She showed that while conservation of momentum follows from homogeneity of the ambient space, conservation of material momentum implies a homogeneous material ``space''.  The inference was that the balance of material momentum follows from variation of the material coordinates, although this process was not explicitly performed.
She constructed several balance laws corresponding to symmetries of the material space and related them to path independent integrals. 
Further applications of the pseudomomentum concept have been extensively explored by Kienzler and G. Herrmann \cite{kienzlerherrmann00}. Maugin was a prolific promoter of these ideas whose viewpoint can be found in several books and review articles \cite{maugin1993, maugin2011, maugin95, maugin2002}. A recent book by O'Reilly \cite{oreilly2017} makes a case for the utility of material momentum balance in the context of rod mechanics.

In the broader context of phase transitions, where the second law of thermodynamics needs to be considered, some researchers including Fried, Gurtin, and Podio-Guidugli  \cite{gurtin00, guidugli2001, friedgurtin2005} consider the balance of configurational momentum to be a basic law of continuum mechanics, on par with the balance of momentum and independent of any constitutive considerations. 
O'Reilly also agrees with this interpretation \cite{oreilly2017}. 
This viewpoint contrasts with that of Rogula, Maugin, Kienzler and G. Herrmann, Rajagopal and Srinivasa \cite{rajagopalsrinivasa2005}, and Yavari and co-workers \cite{yavari06}, who consider the material momentum balance law to be merely the pullback of the spatial momentum balance onto the material's reference configuration. 
Rajagopal and Srinivasa \cite{rajagopalsrinivasa2005} attribute the existence of configurational forces to evolving reference configurations of the body. 
In the present work, which restricts itself to conservative mechanical systems governed by an action principle, both the energy and pseudomomentum balances could be derived from the momentum balance without invoking additional assumptions.  However, instead we employ a variational formulation that is a non-standard extension of the usual action principle of classical mechanics, although it is one not without precedent \cite{edelen1981, MauginTrimarco92mixed, MauginTrimarco92fracture}.

Other notable early works include those of Sturrock on the ``pseudo-energy-pseudo-momentum tensor'' of waves in plasmas \cite{Sturrock62}, Gilbert and Mollow on ``tensor momentum'' associated with elastic vibrations \cite{gilbertmollow1968}, Broer's presentation of the pseudomomentum balance for a moving string \cite{broer70}, Knowles and Sternberg's derivation of material conservation laws for finite elastostatics by application of Noether's theorem to ``translations'' and ``rotations'' of material coordinates \cite{knowlesSternberg1972}, and Fletcher's extension of their results to elastodynamics \cite{fletcher1976}.
Edelen presented a comprehensive treatment of variational elastostatics and its associated conservation laws \cite{edelen1981}.
R. Hill derived the Eshelby tensor for elastostatics using the principle of virtual work \cite{hill1986}. 
Variational derivations of the balance of pseudomomentum may be found in Nelson \cite{nelson91} and Thellung \cite{thellung1994}, in the context of interaction between an electromagnetic field and an elastic dielectric.
The material relabeling symmetry of fluids was exploited by Eckart \cite{eckart1960}, Newcomb \cite{newcomb1967}, Bretherton \cite{bretherton1970}, and Salmon \cite{salmon1988}, and later by M{\"{u}}ller \cite{Muller95} and Padhye and Morrison \cite{padhyemorrison1996}, to derive a variety of known conservation laws.
Benjamin provided a broad commentary on variational principles and conservation laws in fluids \cite{benjamin84}, using the term ``impulse" for Noether's charge in a conservation law arising from variations of independent coordinates.  He did not distinguish between material and spatial coordinates, but remarked that the impulse is in many cases distinct from the momentum.
The term ``impulse'' was inherited by Maddocks and Dichmann in the context of conservation laws in rod dynamics, although they did not invoke a variational principle \cite{maddocksdichman1994}.
Healey used the term ``circulation'' for the the integral of this quantity over a loop of moving string \cite{healey1996}.
The concept of pseudomomentum is used among researchers of disturbance flows in geophysical fluid dynamics.
McIntyre emphasized the distinction between momentum and pseudomomentum of fluid waves as respectively arising from translational invariance of the entire system and of the medium \cite{mcIntyre1981}.
Other examples may be found in the broad works \cite{Shepherd90,BuehlerBOOK} and in many references cited therein.

In this paper, we examine pseudomomentum in a variational setting, deriving bulk and singular balances of momentum, energy, and pseudomomentum from an action principle. The derivation systematically displays the conjugate relationships between the variations and the associated physical quantities.  Just as the balance of momentum can be associated with position, the balances of energy and pseudomomentum can be respectively associated with the independent variables of time and material labels.
After a general discussion, we apply these concepts to various examples, including simple elastic solids, ideal fluids, and thin structures.

We begin in Section \ref{sec:balance_laws} with a procedure for varying a material action simultaneously with respect to material coordinates and time (the independent fields) and the present configuration (the dependent field).  We conduct our calculations in the present configuration, and identify the changes in the Lagrangian density induced by the shifts in the fields.  
A direct consequence of the variational treatment is that the bulk balance equations for material momentum and energy are projections of the momentum balance onto the material tangents and material velocity, respectively.
In Section \ref{sec:first_gradient_theory}, the balance laws for a 
simple elastic field theory are presented, where the Lagrangian density is assumed to be a function of the material velocity and the deformation gradient and to possess explicit dependencies on the independent field variables. 
Through a rearrangement of the pseudomomentum and energy equations, we see that the source terms for these balance laws are the explicit partial derivatives of the Lagrangian density with respect to the reference configuration and time, respectively.
An equivalent derivation in the referential frame is presented in Appendix \ref{app:derivation_reference_coordinates}.  
We invoke Noether's theorem to identify conservation laws arising due to invariance under shifts in the present and reference configurations, and relate the latter to the path independent J-integral.
In Section \ref{sec:ideal_fluids}, we consider material symmetry and the balance laws of an ideal fluid, show that several results pertaining to vorticity, circulation, and helicity can be derived from the general form of the balance of pseudomomentum, and note a connection between circulation and the J-integral. 
In Section \ref{sec:rods_with_variable_cross_sections}, we consider the bulk and singular balance laws for quasistatic \emph{elastica} with inhomogeneous bending stiffness, and apply these to understand the propulsive and reaction forces observed in recent studies of confined rods \cite{cicconofridesimone2015, dalcorso17}. This is perhaps the best illustration of the potential power of the pseudomomentum balance, which provides a simple, almost effortless, derivation of the propulsive ``force'' on the body after a reasonable prescription for singular sources is provided by analogy with the bulk balance law.
We conclude in Section \ref{sec:surfaces} with two examples of conical surface mechanics, for which conserved quantities associated with spatial and material rotational symmetry can be used to classify equilibrium configurations: rotating inertial membranes \cite{Guven13skirts}, and plates with bending energy \cite{guvenmuller2008}.

\section{Balance laws}\label{sec:balance_laws}

We begin by deriving balance laws for momentum, energy, and pseudomomentum from variation of an action simultaneously with respect to dependent and independent variables. Other such treatments exist in the literature \cite{rogula1966, edelen1981, MauginTrimarco92mixed, MauginTrimarco92fracture}, and the required variational machinery can be found in several places \cite{hill1951, rosen1972, barbashov1983, LovelockRund88}.
Our approach differs from these others in several ways. We derive the laws in the current configuration, although all of our fields are defined per unit volume in the reference configuration.
We present the derivation so as to delineate the Noether charges and currents associated with each type of variation.  We also allow for a propagating non-material singular interface in the material, and so obtain both field equations and jump conditions. 

We will consider a body $\mathcal{B}$ as a differentiable, orientable manifold with boundary $\partial\mathcal{B}$.  Physically, this manifold comprises a collection of material elements labeled by attached material coordinates $\mc^i$, where the index $i$ can run over one, two, or three dimensions.
The configuration of the body at time $t$ is an embedding $\bx(\mc^i,t)$ in three-dimensional Euclidean space $\mathds{E}^3$. 
This embedding induces a metric on $\mathcal{B}$.
Many quantities will be defined in a static reference configuration $\rconfig{\bx}(\mc^i)$.
In some situations, it may be necessary or convenient to think of the manifold as Riemannian, carrying its own metric rather than obtaining it from a reference configuration.  Incompatible-elastic bodies have no stress-free reference configuration, and low-dimensional bodies have many such possible configurations. 

We will require \emph{explicit} partial derivatives $\partial_t$ and $\partial_i \equiv \tfrac{\partial}{\partial\mc^i}$, as well as a material time derivative $d_t$, material (noncovariant) derivatives $d_i$, and covariant derivatives $\nabla_i$ and $\rconfig{\nabla}_i$ constructed with the present and reference metrics whose components in the present and reference coordinate bases are, respectively,
 $\g_{ij} = d_i\bx\cdot d_j\bx = \nabla_i\bx\cdot\nabla_j\bx$ and
$\rconfig\g_{ij} = d_i\rconfig\bx\cdot d_j\rconfig\bx = \rconfig\nabla_i\rconfig\bx\cdot\rconfig\nabla_j\rconfig\bx$.  We also define reciprocal bases such that\footnote{Throughout the text the symbol $\delta$ will also be used to denote a small variation in a quantity; no confusion should arise with the index-bearing Kronecker $\delta^i_j$.} $\nabla^i\bx\cdot\nabla_j\bx = \delta^i_j$ and $\rconfig\nabla^i\rconfig\bx\cdot\rconfig\nabla_j\rconfig\bx = \delta^i_j$.

We construct an action using a Lagrangian density $\Lref\left(\mc^i,t;\bx\right)$, where dependence on temporal and material derivatives of $\bx$ is implied, but not explicitly shown for brevity of notation.
The arguments are written such that the dependent fields appearing after the semicolon are considered to be functions of the independent fields before it.  
The density is defined per unit reference volume of the body, which in the examples considered in this paper is equivalent to per unit mass.  However, with the exception of
Appendix \ref{app:derivation_reference_coordinates}, we will work in the present configuration, and thus write the action in the second of the two ways below,
\begin{align}
A = \int_{t_0}^{t_1} \!\! dt \int_{\mathcal{B}} \dvolref\, \Lref (\mc^i, t;\bx) = \int_{t_0}^{t_1} \!\! dt \int_{\mathcal{B}} \dvol\, \JI \Lref (\mc^i, t;\bx)\, .\label{action}
\end{align}
The present and reference volume forms $\dvol=\sqrt{\g}\,d\mc^1 d\mc^2 d\mc^3$ and $\dvolref = \sqrt{\rconfig{\g}}\,d\mc^1 d\mc^2 d\mc^3$ use the metric determinants $\g\equiv \det{\g_{ij}}$ and $\rconfig\g\equiv \det{\rconfig\g_{ij}}$, respectively, and are related by the Jacobian (determinant) $J = \sqrt{\g/\rconfig{\g}}$ such that $\dvol = J \dvolref$.
The description \eqref{action} contrasts with that of geometric energies, such as those describing soap films, in which case it is more natural to work with Lagrangian densities per unit present volume (area).

In the spirit of several prior investigators \cite{rogula1966, edelen1981, herrmannalicia1981,herrmannalicia1982,herrmannalicia1983, MauginTrimarco92mixed, MauginTrimarco92fracture, maugin1993, kienzlerherrmann00}, 
we subject the action \eqref{action} to a set of transformations of both independent and dependent fields: $\mc^i\rightarrow\shift \mc^i$, $t\rightarrow\shift t$, $\bx(\mc^i,t)\rightarrow\shift\bx(\shift\mc^i,\shift t)$. The transformed action is
\begin{align}
\shift{A} = \int_{\shift{t_0}}^{\shift{t_1}} \!\! d\shift{t}\int_{\shift{\mathcal{B}}} \! \shift{\dvol} \, \shift{J}^{-1}\Lref(\shift{\mc}^i, \shift{t}; \shift{\bx})\, .\label{action_transformed}
\end{align}
Shifting the independent variables also transforms the domains of integration in time and space.  
We assume that the transformation involves small shifts of the form
\begin{align}
\shift \mc^i = \mc^i + \delta\mc^i(\mc^j,t)\, ,\qquad \shift t = t +\delta t\, ,\qquad \shift \bx (\shift \mc^i,\shift t) = \bx (\mc^i, t) + \delta\bx (\mc^i,t)\, ,\label{transformations}
\end{align}
where in keeping with a classical treatment, the time shift $\delta t$ is just a uniform constant.
For simplicity, we express the small variations as functions of the original un-transformed independent fields, but these could instead be written as functions of the transformed fields; for example, $\delta\bx(\shift{\mc}^i,\shift{t}) = \delta\bx(\mc^i,t) + \frac{\partial\delta\bx}{\partial\mc^i}\delta\mc^i + \frac{\partial\delta\bx}{\partial t}\delta t + \ldots\,$, with all the terms on the right except the first being of higher order \cite{hill1951}. 

The $\delta$ operator in \eqref{transformations} measures both changes in $\bx$ due to changes in the independent fields, as well as changes in $\bx$ through physical deformation at a \emph{fixed material point}.   Because the two sides of equation \eqref{transformations}$_3$ are functions of two different labels $\mc^i$ and $\shift{\mc}^i$, the $\delta$ operator does not commute with the material derivative.  We thus define \cite{hill1951} an operator $\shop\delta$ which measures changes in the field variable at \emph{a fixed label}, that is, fixed values of $\mc^i$, and therefore commutes with the material derivative,
\begin{align}
\shift\bx(\mc^i,t) = \bx(\mc^i,t) + \shop{\delta}\bx(\mc^i,t)\, .\label{shift_operator}
\end{align}
Using \eqref{transformations}$_3$ and \eqref{shift_operator}, the two variational operators can be related by 
\begin{align}
\delta\bx = \shop{\delta}\bx + \nabla_j\bx\,\delta\mc^j + \dt{\bx}\,\delta t\, .\label{operators_relation}
\end{align}
The second term on the right is the shift in $\bx$ due to shifts in the parameterization  alone.  Note that for a static low-dimensional body such as an elastic surface, any normal variation of the position vector is contained in $\shop{\delta}\bx$.

The change in the action due to the transformation \eqref{transformations} is the difference between \eqref{action} and \eqref{action_transformed},
\begin{align}
\Delta A  = \int_{\shift{t_0}}^{\shift{t_1}} \!\! d\shift{t} \int_{\shift{\mathcal{B}}} \! \shift{\dvol}\, \shift{J}^{-1}\Lref(\shift{\mc}^i, \shift{t}; \shift{\bx}) -  \int_{t_0}^{t_1} \!\! dt \int_{\mathcal{B}} \dvol \, \JI\Lref (\mc^i, t;\bx)\, .\label{delta_A_1}
\end{align}
To evaluate this difference to first order, we manipulate the shifted integral so that it corresponds to the original domain to obtain \cite{noethertavel1971, goldstein2001}
	\begin{align}
	\delta A = \int_{t_0}^{t_1}\!\!dt\,\dt{\left(\delta t \int_{\mathcal{B}}\dvol\JI\Lref\right)} + \int_{t_0}^{t_1}\!\! dt \int_{\mathcal{B}}\dvol\, \nabla_i\left(\JI\Lref\delta\mc^i\right) + \int_{t_0}^{t_1}\!\! dt \int_{\mathcal{B}}\dvol \JI\bdelta\Lref\, .\label{delta_A_B_1}
	\end{align}	
In writing the bulk term on the far right, we note that $\bdelta(\JI\dvol) = \bdelta\dvolref = 0$.  This bulk term contains both Euler-Lagrange content as well as pieces that will contribute to the charge and current on the boundaries after integration by parts.

At this point, several choices are available to us when manipulating the integral.  For the present discussion, we choose to express everything in terms of an integral over the present volume.
Recalling that $\delta t$ is uniform in space, and noting that the material integration limits and the reference volume form $d\bar{V} = \JI dV$ are independent of time, we may rewrite \eqref{delta_A_B_1} simply as
	\begin{align}
	\delta A = \int_{t_0}^{t_1}\!\! dt \int_{\mathcal{B}}\dvol \left[ \JI d_t\left(\Lref\delta t\right) +  \nabla_i\left(\JI\Lref\delta\mc^i\right) + \JI\bdelta\Lref \right] \, .\label{delta_A_B_2}
	\end{align}
This expression separates changes in the action due to shifts in the independent and dependent variables. 
The first two terms account for the shift in the domain of integration in the material coordinates and time, whereas the third term represents the change due to shifts $\shop{\delta}\bx$ in the dependent variable at a \emph{fixed label}.  In terms of formal calculation, the computation of this term involves nothing but the familiar process of variation in which one shifts the dependent fields alone.  
This final term, involving the variation of the Lagrangian density, will generate both bulk and boundary terms through integration by parts, which we write schematically as  
\begin{align}
	\JI\bdelta\Lref = \bm{\mathcal{E}} \cdot\bdelta\bx + \JI d_t \EQ + \nabla_i\EJ^i \, .\label{variation_general}
 \end{align}
 Here $\bm{\mathcal{E}}(\Lref)$ is the Euler-Lagrange operator, and $\EQ$ and $\vEJ$ the temporal and material boundary terms, associated with the Lagrangian density $\Lref$.  The variation of the action \eqref{delta_A_B_2} may be arranged as
\begin{align}
	\delta A &= \int_{t_0}^{t_1}\!\! dt \int_{\mathcal{B}}\dvol \left[ \JI d_t\mathcal{Q} +\nabla\cdot\bm{\mathcal{J}} + \bm{\mathcal{E}}\cdot\bdelta\bx \right] \, ,\label{variation_A} \\
	\mathcal{Q} &= \Lref \delta t + \EQ \, , \\
	\mathcal{J}^i &= \JI \Lref \delta \mc^i + \EJ^i \, .
\end{align}
The form of the Euler-Lagrange term $\bm{\mathcal{E}}$ is in general the same as the boundary terms, but with the possibility of additional source terms.\footnote{While the appearance of $\JI$ may look a bit strange outside of the time derivative, consider that inertial terms are generally proportional to a reference density $\bar{\rho}$ that is independent of time.  Due to conservation of mass, one has $\JI = \sqrt{\bar{g}/g}=\rho/\bar{\rho}$, and moving the reference density through the time derivative, one is left with the present density as coefficient.}
The charge density $\mathcal{Q}$ and current density $\bm{\mathcal{J}}$ are functions of the variations of all the fields.  
The third, Euler-Lagrange term in the integral \eqref{variation_A} delivers the bulk balance laws corresponding to shifts in the dependent and independent variables, in an extension \cite{edelen1981, MauginTrimarco92mixed, MauginTrimarco92fracture} of the usual Hamilton-Lagrange-d'Alembert principle of stationary action. 
When the field variations are symmetries of the system, $\delta A =0$ independently of any considerations regarding the stationarity of $A$, and the essence of Noether's theorem is that the first two terms inside the integral \eqref{variation_A} provide conservation laws\footnote{A strict conservation law form can be seen in a derivation in the referential frame; see for example Appendix \ref{app:derivation_reference_coordinates}.} associated with these symmetries when the Euler-Lagrange term vanishes.

In the following section, we will consider in detail the specific case of a space-filling body with a Lagrangian density dependent on, at most, the first derivative of the position vector.  The classic example of such a 
theory is the elasticity of simple materials \cite{maugin1993}.  Because of its prominence in modern solid mechanics, we will also present a referential version of the above derivation, along with that of Section \ref{sec:first_gradient_theory} below, in Appendix \ref{app:derivation_reference_coordinates}.
 
 Some energies are more naturally formulated with respect to a reference configuration, and others with respect to the present.  An example of the latter would be a geometric energy such as that of a soap film, dependent on the current area of the film and independent of any reference density distribution.
The above framework, discussed with respect to a Lagrangian density $\Lref$ defined with respect to a reference volume, can be augmented with geometric terms $\Lgeom = \JI \Lref = (\rho/\bar{\rho})\Lref$.
For example, an ``inertial soap film" with surface energy $\gamma$ could be represented either by $\Lgeom = \tfrac{1}{2}\rho\dt{\bx}\cdot\dt{\bx} - \gamma$ or $\Lref = \tfrac{1}{2}\rconfig{\rho}\dt{\bx}\cdot\dt{\bx} - J \gamma$.

\section{Simple elastic solid}\label{sec:first_gradient_theory}

In this section, we consider field theories with the form $\Lref(\rconfig{\bx},t;\bx,\dt{\bx},\F)$, where the independent material coordinates are represented by a time-independent reference configuration $\rconfig{\bx}$, and first derivatives of position are represented by a ``deformation gradient'' \cite{gurtin:mechanics} $\F\equiv\rconfig{\nabla}\bx\equiv\frac{d\bx}{d\rconfig{\bx}} = \nabla_i\bx\rconfig{\nabla}^i\rconfig{\bx}$ that applies the gradient $\rconfig{\nabla}$ of the reference space to the position vector $\bx$ of the present configuration (note that $\F$ also serves to transform between bases; for example $d_i\bx\delta \mc^i = \F \cdot d_j \rconfig\bx \delta \mc^j$, and $\Grad () = \grad () \cdot\F$ if () has no free indices).
This description is appropriate, and indeed quite traditional in solid mechanics, for the description of simple elastic bodies that fill some portion of three-dimensional space (that is, $i \in \{1,2,3\}$), but is not suitable for the description of incompatible elastic systems or lower-dimensional bodies such as the elastic surfaces we will consider later in Section \ref{sec:surfaces}.

We will identify the balance and conservation laws that arise from variation of the dependent and independent variables. 
The temporal and material boundary terms from \eqref{variation_general} can be computed as $\EQ = \frac{\partial\Lref}{\partial\dt{\bx}}\cdot\bdelta\bx$ and $\vEJ=\JI\F\cdot\left[\frac{\partial\Lref}{\partial\F}\right]^\mathrm{T}\cdot\,\bdelta\bx$. 
The Euler-Lagrange, charge, and current terms corresponding to the variation \eqref{variation_A} are
\begin{align}
   \bm{\mathcal{E}} &= \JI\frac{\partial\Lref}{\partial\bx} - \JI\dt\left(\frac{\partial\Lref}{\partial\dt{\bx}}\right)
   -\nabla\cdot\left(\JI\F\cdot\left[\frac{\partial\Lref}{\partial\F}\right]^\mathrm{T}\right)\, ,\label{euler_derivative}\\
    \mathcal{Q} &= \left(\Lref\,\delta t + \frac{\partial\Lref}{\partial\dt{\bx}}\cdot\shop{\delta}\bx \right)\, ,\label{noether_charge}\\
   \bm{\mathcal{J}} &= \JI\left(\Lref\,\F\cdot\delta\rconfig{\bx} + \F\cdot\left[\frac{\partial\Lref}{\partial\F}\right]^\mathrm{T}\cdot\shop{\delta}\bx\right)\, .\label{noether_current}
\end{align}
However, to clearly identify the terms power-conjugate to each of the different variational quantities, we rewrite the variations at a \emph{fixed material label} appearing in (\ref{noether_charge}-\ref{noether_current}), to express the charge and current in terms of the total variation $\delta\bx$ at a \emph{fixed material point} as well as variations with respect to the independent material and temporal variables.  Using the relation \eqref{operators_relation} to substitute for $\shop{\delta}\bx$, we rearrange equations (\ref{noether_charge}-\ref{noether_current}) to obtain
\begin{align}
\mathcal{Q} &=\left[\frac{\partial\Lref}{\partial\dt{\bx}}\cdot\delta\bx +\left(\frac{\partial\Lref}{\partial\dt{\bx}}\cdot\F\right)\cdot\left(-\delta\rconfig{\bx}\right) + \left(\frac{\partial\Lref}{\partial\dt{\bx}}\cdot\dt{\bx}-\Lref\right)\left(-\delta t\right)\right]\, ,\label{noether_charge_2}\\
\bm{\mathcal{J}}&=\left[ \left(\JI\F\cdot\left[\frac{\partial\Lref}{\partial\F}\right]^\mathrm{T}\right)\cdot\delta\bx+\left(\JI\F\cdot\left[\frac{\partial\Lref}{\partial\F}\right]^\mathrm{T}\cdot\F - \JI\Lref\,\F\right)\left(-\delta\rconfig{\bx}\right) + \left(\JI\F\cdot\left[\frac{\partial\Lref}{\partial\F}\right]^\mathrm{T}\cdot\dt{\bx}\right)\left(-\delta t\right) \right]\, .\label{noether_current_2}
\end{align}
In this form, we may identify several familiar quantities.
In \eqref{noether_charge_2}, the charges conjugate to $\delta\bx$, $-\delta\rconfig{\bx}$, and $-\delta t$ are, respectively, the (spatial) momentum, pseudomomentum\footnote{A. Golebiewska Herrmann's material momentum \cite{herrmannalicia1982}, or the negative of Peierls's pseudomomentum in equation (2.10) of \cite{peierls1985}.}, and Hamiltonian density \cite{goldstein2001}.  In \eqref{noether_current_2}, the currents conjugate to $\delta\bx$, $-\delta\rconfig{\bx}$, and $-\delta t$ are, respectively, the (spatial) stress, pseudostress\footnote{Material stress, or Eshelby's tensor \cite{eshelby1970}.}, and power expended by the stress. 
As we are working in the present configuration, the stress in question is that of Cauchy  \cite{gurtin:mechanics}, $-\JI\F\cdot\left[\frac{\partial\Lref}{\partial\F}\right]^\mathrm{T}$.  This is the Piola transform of the first Piola-Kirchhoff stress $-\frac{\partial\Lref}{\partial\F}$ whose transpose will appear naturally in the referential frame in Appendix \ref{app:derivation_reference_coordinates}.  Similarly, the Eshelby tensor conjugate to $-\delta\rconfig{\bx}$ in \eqref{noether_current_2} is a transformed version of the usual referential form of this tensor. 

We now proceed to integrate \eqref{variation_A} by parts to obtain bulk balance laws as well as boundary and jump conditions.  
The latter are singular balance laws that hold at an internal non-material surface of discontinuity $\mathcal{S}(t)$.  This surface is assumed to move with some ``velocity'' through the coordinates, whose normal component is denoted $\bar U$.
Let $\uvc{n}$ and $\uvc{N}$ be the unit normals to the external boundary and internal surface of discontinuity.  The relevant forms of the divergence and transport theorems for a piecewise continuous tensorial quantity ${\bf A}$ are \cite{gurtin:mechanics,eringen1980} 
\begin{align}
\int_{\mathcal{B}}\dvol\,\grad\cdot {\bf A}  &= \int_{\partial\mathcal{B}}\!\! \dArea\, \uvc{n}\cdot{\bf A} - \int_{\mathcal{S}(t)}\!\! \dArea\jump{\uvc{N}\cdot{\bf A}}\, ,\label{divergence_theorem}\\
\dt{\int_{\mathcal{B}}\dvol\, \JI{\bf A}} &= \int_{\mathcal{B}}\dvol\,\JI\dt{\bf A} - \int_{\mathcal{S}(t)}\!\! \dArea\jump{\JI U\bf A}\, ,\label{transport_theorem}
\end{align}
where $\jump{\,}$ denotes the jump in the enclosed quantity across the discontinuity.
The quantity $\JI U = \JSI\bar U$, where $\JSI$ is the areal Jacobian at the surface, is in general continuous and so can be moved outside the brackets. 
 In the present consideration of space-filling bodies, $\uvc{N}$ will be continuous as well.
Using (\ref{divergence_theorem}-\ref{transport_theorem}) and \eqref{operators_relation}, we obtain 
\begin{align}
\delta A = \int_{t_0}^{t_1}\!\! dt \int_{\partial\mathcal{B}}\!\!\dArea\,\uvc{n}\cdot\bm{\mathcal{J}}+ \int_{t_0}^{t_1}\!\! dt \int_{\mathcal{S}(t)}\!\! \dArea \jump{-\uvc{N}\cdot\bm{\mathcal{J}} + \JI U\mathcal{Q}} +\int_{t_0}^{t_1}\!\! dt \int_{\mathcal{B}}\dvol \left[\bm{\mathcal{E}}\cdot(\delta\bx - \F\cdot\delta\rconfig{\bx} - \dt{\bx}\,\delta t)\right]\, ,\label{variation_A_final}
\end{align}
where we have used $\delta \bx = \F\cdot \delta\rconfig\bx$. These three integrals provide the (free) boundary conditions, jump conditions, and bulk field equations.  We now consider separately the balance laws conjugate to the variation in the current configuration, material coordinates, and time. 

Pure variations $\delta\bx$ of the current configuration, with $\delta\rconfig{\bx}= \bm{0}$ and $\delta t = 0$, provide the bulk equation, the boundary condition, and the jump condition for momentum,
\begin{align}
\bm{\mathcal{E}}(\Lref) = \JI\frac{\partial\Lref}{\partial\bx} - \JI\dt{\left(\frac{\partial\Lref}{\partial\dt{\bx}}\right)} - \nabla\cdot\left(\JI\F\cdot\left[\frac{\partial\Lref}{\partial\F}\right]^\mathrm{T}\right)&=\bm{0}\quad\text{on}\;\mathcal{B}\, ,\label{physical_momentum_bulk}\\
\uvc{n}\cdot\left(\JI\F\cdot\left[\frac{\partial\Lref}{\partial\F}\right]^\mathrm{T}\right) &= \bm{0}\quad\text{on}\;\partial\mathcal{B}\, ,\label{physical_momentum_boundary}\\
\jump{-\uvc{N}\cdot\left(\JI\F\cdot\left[\frac{\partial\Lref}{\partial\F}\right]^\mathrm{T}\right) + \JI  U \frac{\partial\mathcal{L}}{\partial\dt{\bx}}} &= \bm{0}\quad\text{on}\;\mathcal{S}(t)\, .\label{physical_momentum_jump}
\end{align}
Pure variations $\delta\rconfig{\bx}$ of the reference configuration, with $\delta\bx = \bm{0}$ and $\delta t = 0$, provide balance laws for pseudomomentum,
\begin{align}
\bm{\mathcal{E}}(\Lref)\cdot\F =  \left[\JI\frac{\partial\Lref}{\partial\bx} - \JI\dt{\left(\frac{\partial\Lref}{\partial\dt{\bx}}\right)} - \nabla\cdot\left(\JI\F\cdot\left[\frac{\partial\Lref}{\partial\F}\right]^\mathrm{T}\right)\right]\cdot\F &=\bm{0}\quad\text{on}\;\mathcal{B}\, ,\label{material_momentum_bulk}\\
\uvc{n}\cdot\left(\JI\F\cdot\left[\frac{\partial\Lref}{\partial\F}\right]^\mathrm{T}\cdot\F - \JI\Lref\,\F \right)&=\bm{0}\quad\text{on}\;\partial\mathcal{B}\, ,\label{material_momentum_boundary}\\
\jump{-\uvc{N}\cdot\left(\JI\F\cdot\left[\frac{\partial\Lref}{\partial\F}\right]^\mathrm{T}\cdot\F - \JI\Lref\,\F\right) + \JI  U\left(\frac{\partial\Lref}{\partial\dt{\bx}}\cdot\F\right)}&=\bm{0}\quad\text{on}\;\mathcal{S}(t)\, .\label{material_momentum_jump}
\end{align}
Finally, purely temporal variations $\delta t$, with $\delta\bx = \delta\rconfig{\bx} = \bm{0}$), provide balance laws for energy,
\begin{align}
\bm{\mathcal{E}}(\Lref)\cdot\dt{\bx} =  \left[\JI\frac{\partial\Lref}{\partial\bx} - \JI\dt{\left(\frac{\partial\Lref}{\partial\dt{\bx}}\right)} - \nabla\cdot\left(\JI\F\cdot\left[\frac{\partial\Lref}{\partial\F}\right]^\mathrm{T}\right)\right]\cdot\dt{\bx}&=0\quad\text{on}\;\mathcal{B}\, ,\label{energy_bulk}\\
\uvc{n}\cdot\left(\JI\F\cdot\left[\frac{\partial\Lref}{\partial\F}\right]^\mathrm{T}\cdot\dt{\bx}\right) & =0\quad\text{on}\;\partial\mathcal{B}\, ,\label{energy_boundary}\\
\jump{-\uvc{N}\cdot\left(\JI\F\cdot\left[\frac{\partial\Lref}{\partial\F}\right]^\mathrm{T}\cdot\dt{\bx}\right) + \JI U\left(\frac{\partial\Lref}{\partial\dt{\bx}}\cdot\dt{\bx}-\Lref \right)} &= 0\quad\text{on}\;\mathcal{S}(t)\, .\label{energy_jump}
\end{align}
While the bulk balances for pseudomomentum and energy are simply projections of the momentum balance onto the deformation gradient and velocity, respectively, the boundary and jump conditions are distinct.  Note also that the two sets of vector equations correspond to different ``legs'' of the two-point tensorial quantities they contain; the extant leg in the momentum equations corresponds to the present configuration while that in the pseudomomentum equations corresponds to the reference configuration.
 Referential forms of these balance laws are presented in Appendix \ref{app:derivation_reference_coordinates}.
 Because equations (\ref{material_momentum_bulk}-\ref{material_momentum_jump}) arise from a continuous shift in material coordinates made possible by a continuum description of a body, they have no analogue in a discrete set of particles \cite{herrmannalicia1981,herrmannalicia1983}.

Although the simple relationship between the bulk balance laws in our system implies that satisfaction of the balance of momentum \eqref{physical_momentum_bulk} means that the other balances \eqref{material_momentum_bulk} and \eqref{energy_bulk} hold as well, this obscures a crucial point, namely that the conserved quantities associated with the corresponding symmetries are not identical.
In the following section, we will rearrange these equations to illustrate that the source terms arising from broken spatial, temporal, and material symmetries are mutually independent quantities.

\subsection{Forces and material forces}\label{sec:physical_and_material_forces}

Following A. Golebiewska Herrmann \cite{herrmannalicia1981} and Maugin \cite{maugin1993}, we recast the balances of energy \eqref{energy_bulk} and pseudomomentum \eqref{material_momentum_bulk} into a standard form that clearly reveals the form of the source terms.
Referential expressions that follow a strict conservation law form are presented in Appendix \ref{app:derivation_reference_coordinates}. 

First note that the balance of momentum \eqref{physical_momentum_bulk} can be easily written with a source term on the right hand side,
\begin{align}
 \JI\dt{\left(\frac{\partial\Lref}{\partial\dt{\bx}}\right)} + \nabla\cdot\left(\JI\F\cdot\left[\frac{\partial\Lref}{\partial\F}\right]^\mathrm{T}\right) = \JI\frac{\partial\Lref}{\partial\bx}\, .\label{physical_momentum_balance_law}
\end{align}
Any explicit dependence of the Lagrangian density on the position $\bx$, such as the presence of a gravitational potential, breaks the symmetry of the embedding space and provides a source of momentum.

The balance of energy \eqref{energy_bulk} can be rearranged by employing the chain rule
\begin{align}
\dt{\Lref} = \frac{\partial\Lref}{\partial t} + \frac{\partial\Lref}{\partial\bx}\cdot\dt{\bx} + \frac{\partial\Lref}{\partial\dt{\bx}}\cdot\dt{\dt{\bx}} + \left[\frac{\partial\Lref}{\partial\F}\right]^\mathrm{T}:(\nabla\dt{\bx}\cdot\F)\, ,\label{lagrangian_time_derivative}
\end{align}
where the final term involves the material time derivative of $\bF$; the notation means that the $\nabla$ leg is contracted with the present leg of $\F$, and double contraction associates present and referential legs with their respective counterparts.
After some integration by parts, \eqref{energy_bulk} becomes
\begin{align}
\JI\dt{\left(\frac{\partial\Lref}{\partial\dt{\bx}}\cdot\dt{\bx} - \Lref\right)} + \nabla\cdot\left(\JI\F\cdot\left[\frac{\partial\Lref}{\partial\F}\right]^\mathrm{T}\cdot\dt{\bx}\right) = -\JI\frac{\partial\Lref}{\partial t}\, .\label{energy_bulk_2}
\end{align}
As one might expect, any explicit dependence of the Lagrangian density on the time $t$ manifests as a source term in the energy balance. 

Similarly, the balance of pseudomomentum \eqref{material_momentum_bulk} can be rearranged with the help of the chain rule
\begin{align}
\nabla\Lref\cdot\F = \frac{\partial\Lref}{\partial\rconfig{\bx}} + \frac{\partial\Lref}{\partial\bx}\cdot\F + \frac{\partial\Lref}{\partial\dt{\bx}}\cdot\dt{\F} + \left( \left[\frac{\partial\Lref}{\partial\F}\right]^\mathrm{T}:\nabla\F \right)\cdot\F \, ,\label{lagrangian_covariant_derivative}
\end{align}
where the double contraction involves
both legs (one referential and one present) of $\F$. 
After some integration by parts and use of the Piola identity $\grad\cdot\left(\JI\F\right) = {\bm 0}$, \eqref{material_momentum_bulk} becomes
\begin{align}
\JI\dt{\left(\frac{\partial\Lref}{\partial\dt{\bx}}\cdot\F\right)} + \nabla\cdot\left(\JI\F\cdot\left[\frac{\partial\Lref}{\partial\F}\right]^\mathrm{T}\cdot\F - \JI\Lref\, \F\right) = - \JI\frac{\partial\Lref}{\partial\rconfig{\bx}}\, .  \label{material_momentum_bulk_2_temp}
\end{align}
Any explicit dependence of the Lagrangian density on the reference configuration $\rconfig\bx$ breaks the symmetry of the material continuum and provides a source of pseudomomentum.  
Forms of the balance law \eqref{material_momentum_bulk_2_temp} in present or referential form, with or without the source term, can be found in  \cite{maugin1993,maugin2011,kienzlerherrmann00,thellung1994,rogula1966,knowlesSternberg1972,fletcher1976,herrmannalicia1981,herrmannalicia1982,herrmannalicia1983,cermelliFried1997}.

The source terms in the three balances \eqref{physical_momentum_balance_law}, \eqref{energy_bulk_2}, and \eqref{material_momentum_bulk_2_temp} are entirely independent.
In particular, the balance of pseudomomentum is related to the symmetry of the material continuum, a feature independent of any properties of the embedding space.
Just as the source term in \eqref{physical_momentum_bulk} is often interpreted as a body force, we may interpret the source term in \eqref{material_momentum_bulk} as a ``material body force".
However, it is important to note that these forces are not of the same type.  While spatial (Newtonian) forces are vectors associated with the embedding space, material (Eshelbian) forces are associated with a material space.  Although in the example under consideration, the material space can be thought of as a reference configuration embedded in the same space as the present configuration, with material forces associated with vectors in the reference configuration, and although the present configuration of a space-filling body is often associated with the embedding space itself, this does not mean that spatial and material forces can be conflated or added together in any meaningful way.
They pertain, respectively, to the motion of material bodies in space and the motion of non-material objects within a material.

\subsection{A few symmetries and conservation laws}\label{sec:symmetries_conservation_laws}

Here we apply Noether's theorem to obtain conservation laws\footnote{We use the term ``conservation law'' loosely.} for momentum and pseudomomentum.
We insert variations corresponding to spatial and material symmetries of the action into the general expression
\begin{align}
\JI\dt{\mathcal{Q}} + \nabla\cdot\bm{\mathcal{J}} = 0 \, ,\label{symmetries_general}
\end{align}
where $\mathcal{Q}$ and $\bm{\mathcal{J}}$ are given by \eqref{noether_charge_2} and \eqref{noether_current_2}.  
A static version of this general statement in elasticity can be found in Edelen \cite{edelen1981}.

The embedding space is symmetric under translations $\delta\bx = \bD$ and rotations $\delta\bx = \bD\times\bx$, where $\bD$ is a (small) constant vector.  With $\delta\rconfig{\bx}=\bm{0}$ and $\delta t=0$, we obtain linear and angular momentum conservation laws, 
\begin{align}
\JI\dt{\left(\frac{\partial\Lref}{\partial\dt{\bx}}\right)} + \nabla\cdot\left(\JI\F\cdot\left[\frac{\partial\Lref}{\partial\F}\right]^\mathrm{T}\right)  &= \bm{0}\, ,\label{physical_momentum_conservation}\\
\JI\dt{\left(\bx\times\frac{\partial\Lref}{\partial\dt{\bx}}\right)} + \nabla\cdot\left(\bx\times\JI\F\cdot\left[\frac{\partial\Lref}{\partial\F}\right]^\mathrm{T}\right) &=\bm{0}\, .\label{physical_angular_momentum_conservation}
\end{align}
By comparing \cite{yavari06} the linear momentum conservation law \eqref{physical_momentum_conservation} with the balance law \eqref{physical_momentum_balance_law}, we can see that conservation implies that the Lagrangian density cannot depend explicitly on position $\bx$.

If the material is uniform, ``translations'' in material coordinates $\delta\rconfig{\bx}$ produce the linear pseudomomentum conservation law
\begin{align}
\JI\dt{\left(\frac{\partial\Lref}{\partial\dt{\bx}}\cdot\F\right)} + \nabla\cdot\left( \JI\F\cdot\left[\frac{\partial\Lref}{\partial\F}\right]^\mathrm{T}\cdot\F - \JI\Lref\,\F\right)  &= \bm{0}\, .\label{material_momentum_conservation_2}
\end{align}
In the present context, this means \cite{yavari06} that the Lagrangian density cannot depend explicitly on the reference configuration $\rconfig{\bx}$, as can be seen by comparing \eqref{material_momentum_conservation_2} with \eqref{material_momentum_bulk_2_temp}.
Material ``rotational'' symmetry and angular pseudomomentum conservation will be exploited in Section \ref{sec:surfaces}.  Pseudomomentum conservation laws can be found in \cite{rogula1966,knowlesSternberg1972,fletcher1976,herrmannalicia1981,herrmannalicia1982,herrmannalicia1983, thellung1994, maugin1993, maugin2011,kienzlerherrmann00}.

\subsection{J-integral}\label{sec:fracture_hyperelastic_solids}

As demonstrated by A. Golebiewska Herrmann \cite{herrmannalicia1982,herrmannalicia1983}, conservation laws arising from invariance of material space are intimately related to well-known path independent integrals of hyperelastic fracture mechanics \cite{rice1968,Cherepanov1967}.  Consider the conservation of ``translational'' material momentum \eqref{material_momentum_conservation_2} integrated over an arbitrary volume $V$ with boundary $\partial V$ and unit normal $\uvc{\nu}$, 
\begin{align}
\int_{V}\!\dvol\left[ \JI\dt\left(\frac{\partial\Lref}{\partial\dt{\bx}}\cdot\F\right) +\nabla\cdot\left(\JI\F\cdot\left[\frac{\partial\Lref}{\partial\F}\right]^\mathrm{T}\cdot\F - \JI\Lref\,\F\right) \right] = \bm{0}\, .\label{material_momentum_bulk_solid_integrated}
\end{align}
If $V$ encapsulates a defect such as an inclusion or crack tip-- a point where the conservation law fails to hold-- the right hand side of the above equation need not be zero.  Such a source term would represent the total material force on the defect that seeks to drive it through the material rather than through space.  Equation \eqref{material_momentum_bulk_solid_integrated} 
 is known as the dynamic generalization of the J-integral \cite{markenscoff2006,maugin95,nakamurashish1985}. 
Markenscoff \cite{markenscoff2006} also discusses the corresponding ``rotational'' L-integral. For the static case, the time derivative vanishes and the divergence term may be written as a surface integral
\begin{align}
\int_{\partial V}\!\! \dArea\,\uvc{\nu}\cdot\left(\JI\F\cdot\left[\frac{\partial\Lref}{\partial\F}\right]^\mathrm{T}\cdot\F - \JI\Lref\,\F\right) = \bm{0}\, ,\label{material_momentum_surface_integral}
\end{align}
the original J-integral of Rice \cite{rice1968} and Cherepanov \cite{Cherepanov1967}.

\section{Ideal Fluid}\label{sec:ideal_fluids}

In this section, we derive the balances of momentum, pseudomomentum, and energy for an inviscid, incompressible fluid in the framework of Section \ref{sec:balance_laws}.  
We also demonstrate that the conservation of several important quantities, namely vorticity, circulation, and helicity, can be seen as a consequence of pseudomomentum balance and material symmetry.  
Many authors have discussed the role of symmetry \cite{eckart1960,newcomb1967,bretherton1970,salmon1988,Muller95,padhyemorrison1996}, but we are unaware of a prior demonstration that a single local balance law implies conservation of all of these quantities, some of which are defined as integrals.

Variational derivations of both Lagrangian and Eulerian inviscid fluid equations exist \cite{eckart1960,seligerwhitham1968,bretherton1970,salmon1988}.
As fluid mechanics is often considered from an Eulerian point of view, the utility of material symmetry may not be immediately obvious. However, many classical results in fluid mechanics are of a material character, such as Kelvin's circulation theorem describing the conservation of ideal fluid impulse evaluated over a material loop.

Rather than a reference configuration in the sense of an elastic solid, the fluid will be given an arbitrarily chosen set of material labels that convect with the flow.  We still define a Lagrangian density in terms of the density $\rconfig\rho$ at some reference state (presumed uniform for simplicity), although as the fluid is incompressible the distinction between this and the present density $\rho$ affects only formal manipulations.  
In terms of the present configuration of a fluid $\bx\equiv\bx(\mc^i,t)$, 
\begin{align}
A =\int_{t_0}^{t_1}\!\! dt \int_{\mathcal{B}} \dvol \,\JI\left[\tfrac{1}{2}\rconfig{\rho}\dt{\bx}\cdot\dt{\bx} + p\left(J - 1 \right)\right]\, ,\label{ideal_fluid_action}
\end{align}
where the pressure $p$ is a Lagrange multiplier enforcing the incompressibility constraint $J=1$.
We employ $\shop \delta J = J\nabla^i\bx\cdot\nabla_i\shop \delta \bx$ and subsequently invoke $\JI \rconfig{\rho} = \rho$ and $J=1$ to obtain 
\begin{align}
\delta A = \int_{t_0}^{t_1}\!\! dt \int_{\mathcal{B}} \dvol\, \bigg(&\dt{\bigg[\rho\dt{\bx}\cdot\delta\bx + \rho\dt{\bx}\cdot\nabla_j\bx\left(-\delta\mc^j\right) + \tfrac{1}{2}\rho\dt{\bx}\cdot\dt{\bx}(-\delta t)\bigg]}\nonumber\\
& + \nabla_i\bigg[ p\nabla^i\bx\cdot\delta\bx + \left(p - \tfrac{1}{2}{\rho}\dt{\bx}\cdot\dt{\bx}\right)\left(-\delta\mc^i\right) + p\nabla^i\bx\cdot\dt{\bx}(-\delta t) \bigg]\nonumber\\
& +  \bigg[- \dt{\left(\rho\dt{\bx}\right)} - \nabla_i\left( p\nabla^i\bx\right)\bigg]\cdot\left(\delta\bx - \nabla_j\bx\,\delta\mc^j - \dt{\bx}\,\delta t\right)\bigg)\, .\label{ideal_fluid_action_variation}
\end{align} 
The bulk equations for momentum, pseudomomentum, and energy are thus
\begin{align}
\rho d_t^2\bx + \nabla_i\left(p\nabla^i\bx\right) &= \bm{0}\, ,\label{ideal_fluid_physical_momentum_bulk}\\
\dt{\left(\rho\dt{\bx}\cdot\nabla_i\bx\right)} + \nabla_i\left(  p - \tfrac{1}{2}\rho\dt{\bx}\cdot\dt{\bx}\right) &=0\, ,\label{ideal_fluid_material_momentum_bulk}\\
\dt{\left(\tfrac{1}{2}\rho\dt{\bx}\cdot\dt{\bx}\right)} + \nabla_i\left(p\nabla^i\bx\cdot\dt{\bx}\right) &= 0\, .\label{bernoulli_1}
\end{align}
Note that because of the way we have written the dependencies of the action, the pseudomomentum balance \eqref{ideal_fluid_material_momentum_bulk} is obtained in component form.  The pseudomomentum is the quantity whose components are inside the time derivative; it is also known as the impulse \cite{benjamin84} or ``vortex momentum" \cite{kuzmin1983}. 
The pseudomomentum and energy balances \eqref{ideal_fluid_material_momentum_bulk} and \eqref{bernoulli_1} are rearrangements, by a chain rule procedure akin to that in Section \ref{sec:physical_and_material_forces}, of projections of the momentum balance \eqref{ideal_fluid_physical_momentum_bulk} onto the tangents $\nabla_i\bx$ and velocity $\dt\bx$, respectively.

Terms such as $\nabla^2 \bx$ appear in the momentum  \eqref{ideal_fluid_physical_momentum_bulk} and energy  \eqref{bernoulli_1}  equations.  If $\bx$ were a surface, these would represent normal vectors, but if $\bx$ is merely a space-filling blob of fluid in flat space, these terms vanish, giving for example $\rho d_t^2\bx + \nabla p = \bm{0}$ for the momentum equation.
For the energy equation, it is more useful to note that the flow is incompressible ($\nabla_i v^i = 0$, where $\bv \equiv \dt{\bx}$), giving $d_t\left(\tfrac{1}{2}\rho\bv\cdot\bv\right) + v^i\nabla_ip=0$. Noting that $d_t = \frac{\partial }{\partial t} + v^i\nabla_i $, we can write 
\begin{align}
\dt{\left(\frac{1}{2}\rho\bv\cdot\bv + p \right)} = \frac{\partial p}{\partial t}\, ,\label{bernoulli_generalized_lagrangian}
\end{align}
a form of Bernoulli's equation equivalent to Eckart's (3.14) \cite{eckart1960}.  Alternately, keeping in mind that $\rho$ is uniform and constant, we can write a more familiar expression,
\begin{align}
	\bv\cdot\left[ \frac{\partial\bv}{\partial t} + \nabla\left(\frac{1}{2}\bv\cdot\bv + \frac{p}{\rho} \right) \right] &=0 \, ,	\label{bernoulli_generalized_eulerian}
\end{align}
involving the streamline derivative $\bv\cdot\nabla$.  

The pseudomomentum and energy equations are projections of the momentum equation onto the tangents $d_i \bx = \nabla_i\bx$ and velocity $d_t \bx$, respectively.  One consequence is that when the flow can be expressed as a steady velocity field, the streamline component of the pseudomomentum equation expresses the same content as the energy equation. This is partly why one of the present authors misleadingly identified the conserved quantity associated with the material symmetry of a flowing string with Bernoulli's constant in \cite{HannaPendar16}.

Modifications to an ideal fluid might lead to source terms in any of the balance laws (\ref{ideal_fluid_physical_momentum_bulk}-\ref{bernoulli_1}).  We note that if the resulting source term in the pseudomomentum balance \eqref{ideal_fluid_material_momentum_bulk} takes the form of a gradient of a scalar, the quantities discussed in the following three subsections will still be conserved.

\subsection{Vorticity}

The vorticity is defined as the curl of the velocity, $\omega^i \equiv \epsilon^{ijk}\nabla_j v_k$.  Applying the curl\footnote{We note that $\epsilon^{ijk}\nabla_j = \epsilon^{ijk}d_j$ and, for incompressible flow, $\sqrt{g} = \sqrt{\bar g}$, so $\epsilon^{ijk}\nabla_j = \bar\epsilon^{ijk}\bar\nabla_j$.} to the pseudomomentum equation \eqref{ideal_fluid_material_momentum_bulk} and noting that for constant $\rho$, the time derivative commutes with the metric determinant implicit in the alternating tensor, we obtain
\begin{align}
\rho\dt\omega^i=0\, .\label{ideal_fluid_cauchys_invariants}
\end{align}
The vorticity equation \eqref{ideal_fluid_cauchys_invariants} takes this simple form because the contravariant components, in material coordinates, of any field convecting with (``frozen in'') the flow are such that their material time derivative vanishes \cite{thiffeault2001}.
These components of vorticity may be identified with Cauchy's invariants \cite{caseynaghdi1991,frischvillone2014}.  


Integrating the pseudomomentum equation \eqref{ideal_fluid_material_momentum_bulk} with respect to time, and noting that $\nabla_i$ is equivalent to $d_i$ when acting on a scalar, so can be interchanged with time derivatives and integrals, we obtain the Cauchy-Weber integral relation \cite{lamb1945,bennett2006},
\begin{align}
\rho \left(v_i - v_i |_{t=0}\right)+ d_i\left[\int_0^t\!\! dt \left(p - \tfrac{1}{2}\rho\bv\cdot\bv\right) \right] = 0 \, .\label{ideal_fluid_webers_integral}
\end{align}

\subsection{Circulation}

Integrating the component of the pseudomomentum equation \eqref{ideal_fluid_material_momentum_bulk} directed along a closed material loop ($i=l$), $\oint d\mc^l $, and noting the equivalence of $\nabla_l$ and $d_l$ when acting on a scalar, the time-independence of $d\mc^l$ and $\rho$, and the uniformity of $\rho$, we obtain Kelvin's circulation theorem,
\begin{align}
	\rho d_t\left[\oint \dt{\bx}\cdot d\bm{l}\right]  =0\, . \label{kelvin_circulation_theorem}
\end{align} 
The connection between this theorem and material symmetry, also known as ``relabeling symmetry" or ``exchange invariance", is discussed in several works \cite{eckart1960,bretherton1970,newcomb1967,salmon1988,Muller95,padhyemorrison1996}.

In two dimensions, the integration performed here is analogous to the derivation of the J-integral in Section \ref{sec:fracture_hyperelastic_solids}.  This analogy between solids and fluids has been noted by Cherepanov \cite{cherepanov1977}, A. Golebiewska Herrmann \cite{herrmannalicia1983}, Atilgan \cite{atilgan1997}, and Maugin \cite{maugin2011}.  
Were material symmetry to be broken, for example by the presence of a body inside the loop, the integral could be nonzero, indicating the presence of a material force driving the body through the fluid.

\subsection{Helicity}

We apply Noether's theorem\footnote{We can instead initiate this derivation by ``contracting'' the components of the pseudomomentum equation \eqref{ideal_fluid_material_momentum_bulk} with those of the vorticity $\omega^i=\bar\omega^i$, and using the latter's properties to move them inside the time derivative and divergence.}, using the pseudomomentum conservation law associated with a general symmetry in material coordinates,
\begin{align}
\dt{\left(\rho v_i\delta\mc^i\right)} + \nabla_i\left[\left(p - \tfrac{1}{2}\rho\bv\cdot\bv\right)\delta\mc^i\right] = 0\, ,\label{fluid_material_conservation_law_general}
\end{align}
and consider a coordinate shift that follows the vorticity field, $\delta\mc^i = \epsilon\omega^i$, where $\epsilon$ is a small constant.  
This shift has the property that $\nabla_j\left(\epsilon\omega^j\right)=0$ \cite{Muller95, padhyemorrison1996}.
Integrating over a material volume, noting that $\rho\dvol$ is time-independent, and applying the divergence theorem,
\begin{align}
\dt \int_{V}\!\dvol\, \rho v_i \omega^i + \int_{\partial V}\!\! \dArea\,n_i\left(p - \tfrac{1}{2}\rho\bv\cdot\bv\right)\omega^i  = 0\, .
\end{align}
If the surface $\partial V$ is such that $n_i\omega^i=0$, the helicity within the volume $V$ is conserved \cite{moffatt1969, yahalom1995}, 
\begin{align}
\rho\dt\left[\int_V\!\dvol\, v_i \omega^i \right] = 0\, ,\label{fluid_conservation_of_helicity}
\end{align}
where the final manipulation uses the uniformity of $\rho$ and the individual time-independence of $\rho$ and $\dvol$.

\section{Non-uniform \emph{elastica}}\label{sec:rods_with_variable_cross_sections}

Elastic beams or rods whose properties vary along their length provide excellent one-dimensional demonstrations of concepts related to pseudomomentum. 
One such property is a variable bending stiffness arising from changes in cross section.
An early theoretical investigation is that of Kienzler and G. Herrmann, who considered discontinuities in stiffness in a beam \cite{kienzlerherrmann1986}.
More recently, Bigoni and co-workers have performed a very interesting series of experiments involving both continuous and discontinuous variation in the bending and torsional stiffness of rods as well as confinement conditions imposed on these rods \cite{bigoni15, bigoni14, dalcorso17}.
This group has also offered a theoretical analysis that, we believe, incorrectly conflates forces and material forces, lumping them both under a general heading of configurational or ``Eshelby-like'' effects.
In this section, we present our perspective, which is heavily influenced by the analyses of Kienzler and G. Herrmann \cite{kienzlerherrmann1986} and O'Reilly \cite{oreilly07,oreilly2017,oreilly2015eshelby}, and partially laid out in prior publications \cite{Hanna18, SinghHanna19}.  After presenting the bulk and singular balance laws for momentum and pseudomomentum for a non-uniform planar Euler \emph{elastica}, we apply our approach specifically to the problem of planar serpentine locomotion of a rod through a curved channel \cite{cicconofridesimone2015, dalcorso17}, and attempt to delineate which forces appearing in the problem are actual forces and which are configurational forces.
We demonstrate that the propulsive material force on the confined rod can be obtained directly by integrating the pseudomomentum balance, and obtain reaction forces at points of geometric and material discontinuity from the singular pseudomomentum balance without any appeal to micromechanical arguments \cite{dalcorso17}.  
As in O'Reilly's analysis \cite{oreilly2015eshelby, Hanna18} of Bigoni's sleeve constraint \cite{bigoni15}, this approach requires a prescription for singular sources of pseudomomentum, but once this leap has been taken the results follow immediately.

To facilitate comparison we adopt, to the extent possible, notation from prior works referred to in this section.

\subsection{Balance laws for planar \emph{elastica}}

We consider a static planar configuration of an inextensible elastic curve $\bx(s)$, where $s$ is both arc length and a material coordinate.
A relevant action for an \emph{elastica} with position-dependent stiffness $B(s)$ is
\begin{align}
A = \int_{s_1}^{s_2}\!\! ds \left(-\tfrac{1}{2}\right)\left[B(s)\bOmega\cdot\bOmega + \sigma\left(\ds{\bx}\cdot\ds{\bx} - 1\right)\right]\, ,\label{elastica_action}
\end{align}
where $\bOmega \equiv \ds{\bx}\times d_s^2\bx$ is the Darboux vector, pointing out of the plane, whose squared magnitude is the square of the rod curvature $\kappa$.
We write $B(s)$ to emphasize that the bending stiffness varies along the length of the rod; all other quantities appearing in the brackets, including the Lagrange multiplier $\sigma$, are also functions of $s$.
 Since the rod is inextensible, comparison can be made to the general approach by noting that the volume form is $1ds$ and $J=1$.
 
The variation of the action may be written \cite{SinghHanna19} in terms of the contact force $\bn$, the contact moment $\bmo$, and the only component of the material stress $c$,
\begin{align}
&\delta A = \int_{s_1}^{s_2} \!\! ds \left[\ds{\left( - \bn\cdot\delta\bx - \bmo\cdot\delta\bOmega +c\,\delta s\right)} + \ds{\bn}\cdot\left(\delta\bx - \ds{\bx}\,\delta s\right)\right] \, ,\label{elastica_action_variation} \\
&\bn = \sigma\ds{\bx} - \ds{\left(B(s)d_s^2\bx\right)}\, ,\quad
\bmo = B(s)\bOmega\, ,\quad
c =  \bn\cdot\ds{\bx} + \bmo\cdot\bOmega - \tfrac{1}{2} B(s)\bOmega\cdot\bOmega\, .\label{elastica_force_moment_material_force_definition}
\end{align}
Furthermore, assuming a single point of discontinuity at $s=s_0$ and applying the divergence theorem for piecewise continuous fields, 
\begin{align}
\delta A = \left(- \bn\cdot\delta\bx - \bmo\cdot\delta\bOmega +c\,\delta s \right) |_{s_1}^{s_2} + \jump{ \bn\cdot\delta\bx + \bmo\cdot\delta\bOmega -c\,\delta s }|_{s=s_0} + \int_{s_1}^{s_2}\!\! ds \left[\ds{\bn}\cdot\left(\delta\bx - \ds{\bx}\,\delta s\right)\right]\, ,\label{elastica_action_variation_integrated}
\end{align}
from which expression we may directly obtain boundary conditions, jump conditions, and bulk field equations.  Momentum balance is given by those terms conjugate to $\delta \bx$ with $\delta s = 0$, and pseudomomentum by those conjugate to $\delta s$ with $\delta \bx = \bm{0}$.

The bulk momentum and pseudomomentum balances are 
\begin{align}
 \ds{\bn} &= {\bm 0} \, , \label{elastica_conservation_linear_momentum}\\
 \ds{c} & = -\tfrac{1}{2}\partial_s B(s)\kappa^2 \, . \label{elastica_material_momentum_bulk}
\end{align}
The pseudomomentum balance $\ds{\bn}\cdot\ds{\bx} = 0$ is rearranged into the form \eqref{elastica_material_momentum_bulk} by a chain rule procedure akin to that in Section \ref{sec:physical_and_material_forces}.  The explicit dependence of the action on the coordinate $s$ breaks the material symmetry of the rod, giving rise to the source term on the right hand side.

The corresponding singular balances at the point of discontinuity $s_0$ are
\begin{align}
\bR + \jump{\bn} &= \bm{0}\, ,\label{elastica_physical_force_jump}\\
Y + \jump{c} &= 0\, ,\label{elastica_material_force_jump}
\end{align}
with allowance for singular supplies of momentum $\bR$ and pseudomomentum $Y$ that do not explicitly appear in the action \cite{Hanna18}.  The quantity $\bR$ represents such things as reaction forces from external constraints at the discontinuity, as might arise at the edge of a sleeve \cite{bigoni15}.  The interpretation of $Y$ is still a matter of discussion in the literature \cite{oreilly07,oreilly2017, Hanna18} and will be considered for our specific problem in the following section.

\subsection{Spatial and material sources}\label{sec:sources}

While the identification of the momentum source $\bR$ is straightforward, we lack a general prescription for the pseudomomentum source $Y$.  There is currently no generally agreed upon conceptual framework that provides a physical understanding of what a source of pseudomomentum means.
The prescription of the pseudomomentum source term at a point of discontinuity was inferred from the bulk balance law for a simple transversely loaded elastic beam with continuous curvature  
by Kienzler and G. Herrmann \cite{kienzlerherrmann1986}. 
O'Reilly \cite{oreilly07} has argued that, in general, the source term $Y$ is a constitutive parameter related to the power input $\tilde{E}$ across a moving discontinuity $s_0(t)$ through
$Y\dot{s}_0 = \tilde{E} - \bR\cdot\bv_0 - \bM\cdot\bomega_0$, where $\dot{s}_0$ is the ``velocity'' of the discontinuity through the body, $\bR$ and $\bM$ are a point force and moment acting at the discontinuity, and $\bv_0$ and $\bomega_0$ are the velocity and angular velocity of the spatial point associated with the discontinuity.\footnote{For a different perspective that would lead to the same conclusions, see \cite{Hanna18}.} 
The particular class of problems we are considering consists of the motion of an elastic rod through a space-fixed, frictionless channel.
There are discontinuities in the constraint--- for example, the curvature of the channel--- as well as in the properties of the rod itself.  The former type of discontinuity is fixed in space ($\bv_0 = \bomega_0=\bm 0$), and as no net power is input or dissipated, $\tilde{E}=0$ and thus $Y$ also vanishes.  
For the latter type, we might seek insight from the balance of energy across the discontinuity.  Instead, however, we propose to infer the prescription of $Y$ for any discontinuity from the source term in the continuous balance law for material momentum \eqref{elastica_material_momentum_bulk}, leading to 
\begin{align}
Y &= 0 \qquad\qquad\qquad\,\, \text{for discontinuities in the channel properties}\, ,\label{elastica_channel_source}\\
Y &= \tfrac{1}{2}\jump{B(s)}\kappa^2 \qquad \text{for discontinuities in the material properties}\, ,\label{elastica_material_source}
\end{align}
where clearly \eqref{elastica_channel_source} is just a specific case of \eqref{elastica_material_source}.
The prescription \eqref{elastica_channel_source} is consistent with O'Reilly \cite{oreilly07} and with our prior work \cite{Hanna18}, while the prescription \eqref{elastica_material_source} is consistent with Kienzler and G. Herrmann \cite{kienzlerherrmann1986}.

Thus we make a distinction between broken symmetries in the environment and in the material itself, with only the latter giving rise to a material force.  In the original problem considered by Bigoni and co-workers \cite{bigoni15}, a uniform rod slides in and out of a frictionless sleeve constraint. The relevant discontinuity is the edge of the sleeve, and the prescription $Y=0$ directly provides an interesting relationship between reaction forces and moments at this point \cite{oreilly2015eshelby, Hanna18}.  Although it is quite useful to examine this problem from the viewpoint of configurational balance, it is misleading to invoke a ``configurational force'' in explanations of the rod's behavior, as any such pseudoforce is zero.

Armed with this prescription, we proceed in the next section to analyze the problem of ``serpentine locomotion" \cite{cicconofridesimone2015, dalcorso17} of an elastic rod through a curved, frictionless channel.

\subsection{Serpentine locomotion}\label{sec:serpentine_locomotion}

\begin{figure}[h!]
	\vspace{-0.5cm}
	\includegraphics[width=10cm]{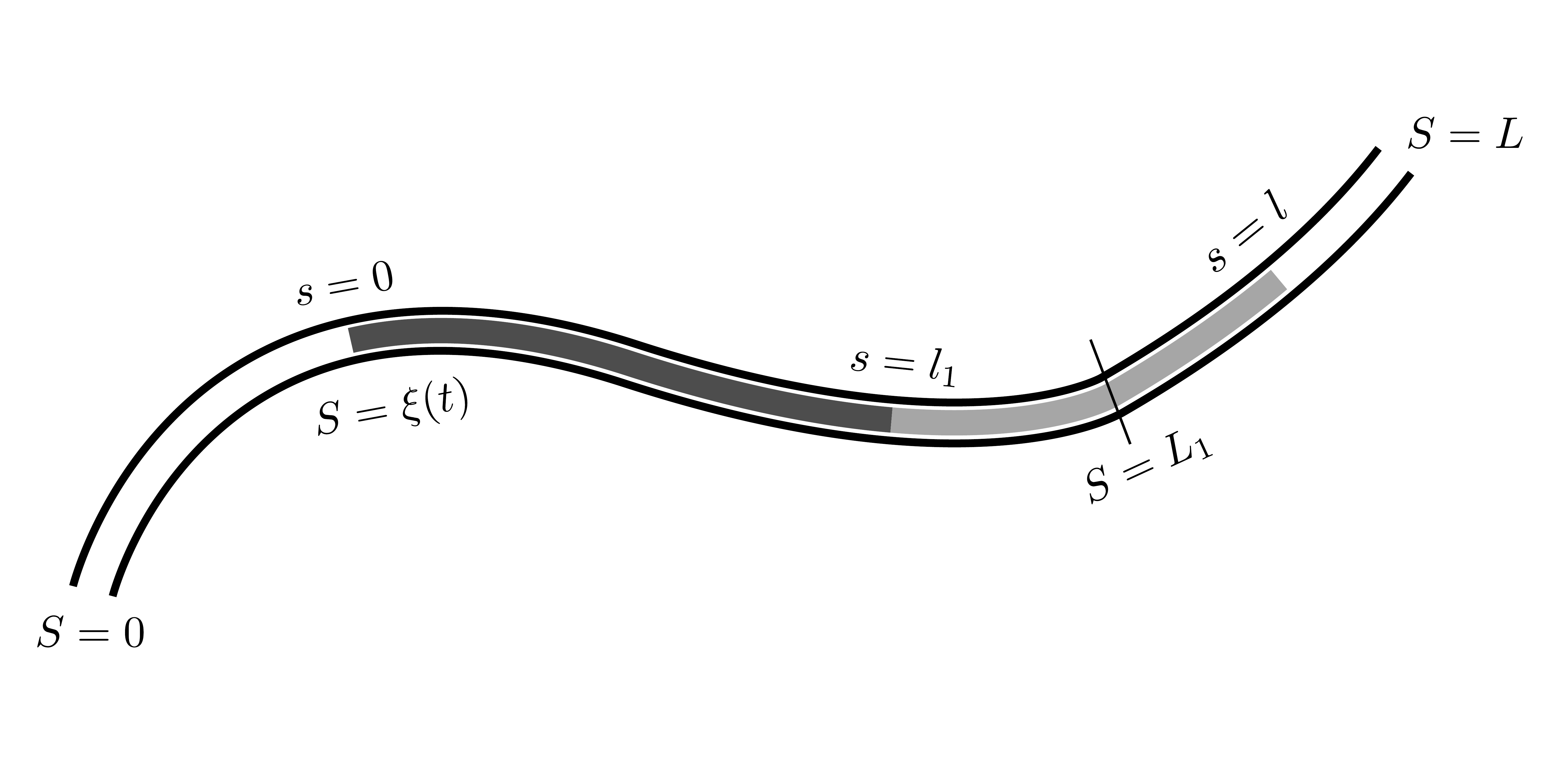}
	\captionsetup{margin=2cm}
	\vspace{-0.75cm}
	\caption{After \cite{dalcorso17}. A planar rod of length $l$ moving in a frictionless channel of length $L$. Here $s$ and $S$ denote the arc length values measured from the left end of the rod and the channel, respectively, with the left end of the rod located at $S=\xi(t)$. 
The thin lines indicate a discontinuity in channel curvature at $S=L_1$, and the two colors in the rod indicate a discontinuity in stiffness $s=l_1$.}
	\label{Fig:sleeve}
\end{figure}

The passive motion of a variable-property rod through a variable-curvature channel, as presented by previous authors \cite{cicconofridesimone2015, dalcorso17}, is portrayed in Figure \ref{Fig:sleeve}.  
 A planar rod of length $l$ moves in a frictionless channel of length $L$.
The coordinates $s$ and $S$ are the arc length values measured from the left ends of the rod and the channel, respectively.   In terms of the channel coordinate, the left end of the rod is located at a time-dependent location $S=\xi(t)$. 
There is a discontinuity in channel curvature $\chi(S)$ at the point $S=L_1$, and a discontinuity in rod stiffness $B(s)$ at the point $s=l_1$.
 For simplicity, we consider only these single discontinuities, the curvature and stiffness being uniform elsewhere.  Continuous variation in these properties can be treated straightforwardly using bulk balance laws.
 The rod is fully constrained so that its curvature $\kappa(s,t) = \chi(\xi(t) + s )$ matches that of the channel.
 The rest curvature of the rod is zero, in contrast to the variable rest curvature considered in \cite{cicconofridesimone2015}.

If we allow for motion of the rod, the static pseudomomentum balance \eqref{elastica_material_momentum_bulk} will be augmented by an inertia term, namely the tangential projection which, in the present constrained case, is the only component of the inertia.
Integrating the static equation \eqref{elastica_material_momentum_bulk} over the rod provides the unbalanced ``propulsive force''  \cite{cicconofridesimone2015, dalcorso17} that tends to move the rod through the channel:
\begin{align}
P(t) &= \int_0^l\!\! ds\,\left[\ds{c} + \tfrac{1}{2}\partial_s B(s)\kappa^2\right] \, , \nonumber \\
	   &=-c(0,t) + c(l,t) - \jump{c(l_1,t)} + \int_0^{l_1^-}\!\! ds\, \tfrac{1}{2}\partial_s B(s)\kappa^2 + \int_{l_1^+}^{l}\!\! ds\, \tfrac{1}{2}\partial_s B(s)\kappa^2\, .\label{elastica_propulsive_force_1}
\end{align}
In this expression, we have left out a jump term at $S=L_1$, because we anticipate that $c$ will be continuous there by way of the prescription \eqref{elastica_channel_source}.  There will be no contribution to the propulsive material force from discontinuities in the imposed constraints.
Using the definition \eqref{elastica_force_moment_material_force_definition}, singular balance \eqref{elastica_material_force_jump}, and prescription \eqref{elastica_material_source}, we can evaluate \eqref{elastica_propulsive_force_1} to be
\begin{align}
P(t) = \tfrac{1}{2}B(0)\kappa^2(0,t) - \tfrac{1}{2}B(l)\kappa^2(l,t) + \tfrac{1}{2}\jump{B(l_1)}\kappa^2(l_1,t) + \int_0^{l_1^-}\!\! ds \, \tfrac{1}{2}\partial_s B(s)\kappa^2(s,t) +  \int_{l_1^+}^{l}\!\! ds \, \tfrac{1}{2}\partial_s B(s)\kappa^2(s,t) \, ,\label{elastica_propulsive_force_final}
\end{align}
in agreement with equation (2.14) of \cite{dalcorso17}.

The first term in the pseudomomentum \eqref{elastica_force_moment_material_force_definition} is the tangential component of the contact force.  Thus, we may also obtain the tangential reaction forces at the rod ends and at the points of discontinuity in the rod stiffness and the channel curvature directly from this definition, the jump condition \eqref{elastica_material_force_jump}, and the prescription (\ref{elastica_channel_source}-\ref{elastica_material_source}):
\begin{align}
 \bn\cdot\ds{\bx} |_{0} &= -B(0)\kappa^2(0,t)\, ,\label{elastica_axial_force_jump_0}\\
 \bn\cdot\ds{\bx} |_{l} &= -B(l)\kappa^2(l,t)\, ,\label{elastica_axial_force_jump_l}\\
 \jump{\bn\cdot\ds{\bx}} |_{l_1} &= -\jump{B(l_1)}\kappa^2(l_1,t)\, ,\label{elastica_axial_force_jump_l1}\\
 \jump{\bn\cdot\ds{\bx}} |_{L_1} &= - \tfrac{1}{2}B(L_1 - \xi)\jump{\chi^2(L_1)} \, .\label{elastica_axial_force_jump_L1}
\end{align}
These relations are equations (2.27-2.30) of \cite{dalcorso17}, where they were computed by micromechanical arguments.

\section{Isometric deformations of conical sheets}\label{sec:surfaces}

We conclude our set of examples with two problems in the mechanics of thin sheets, modeled as flexible, inextensible surfaces, with conical metrics.
As was already implicit in our treatment of elastic rods in Section \ref{sec:rods_with_variable_cross_sections}, these bodies are best considered as Riemannian manifolds endowed with a low-dimensional reference metric, as there is no meaningful distinction between a continuum of possible reference configurations that might correspond to a strain-free embedding of this metric.
The action is written in terms of an embedding $\bx(\mc^\alpha, t)$, $\alpha \in \{1,2\}$, in $\mathds{E}^3$, and a reference metric $\bar{a}_{\alpha\beta} (\mc^\gamma)$. 
In these particular examples, the actual metric adopted $a_{\alpha\beta} = \nabla_\alpha\bx\cdot\nabla_\beta\bx$ will be identical to the reference metric, as enforced by a Lagrange multiplier $\sigma^{\alpha\beta}$.

These systems are characterized by a rotational symmetry of the embedding space, as well as a ``rotational'' material symmetry of the surface, each corresponding to a conserved quantity.  However, the embedding itself will be a generalized cone, not spatially rotationally symmetric except in special cases.  
While our analysis in terms of pseudomomentum is new, much of the structure and results of this section are a result of prior work, in particular the formalism of Guven and co-workers \cite{Guven13skirts, guvenmuller2008}.  One difference is that our actions are written with respect to a reference area, but as all deformations are isometric this has only a formal significance in redefining the multiplier associated with the metric.  

To facilitate comparison we adopt, to the extent possible, notation from prior works referred to in this section.  To avoid confusion with other quantities denoted by a letter $J$, the areal inverse Jacobian is written using the metric determinants explicitly.

The surfaces we consider are conical embeddings $\bx(r,s,t)=r \uvc{u}(s,t)$ parameterized by radial and circumferential material coordinates $r$ and $s$ and the time $t$.  The two tangents to the surface are $d_r\bx = \uvc{u}$ and $d_s\bx = r\uvc{t}$, and the only nonzero components of the metric are $a_{rr}=1$ and $a_{ss}=r^2$.  The normal is $\uvc{u}\times\uvc{t}=\uvc{n}$.
The following relations describe the rotation of a surface-adapted orthonormal frame \cite{guvenmuller2008} along the circumferential coordinate $s$, 
\begin{align}
d_s\uvc{u} = \uvc{t}\, ,\quad d_s\uvc{t} = -\uvc{u} - k\uvc{n}\, ,\quad d_s\uvc{n} = k\uvc{t} \, , \label{cone_relations}
\end{align}
where $k(s,t)$ is a measure of curvature that does not depend on the radial coordinate.
The coordinates $r$ and $s$ are attached to the sheet, and as such are entirely distinct from spatial cylindrical coordinates that might be also used to describe any embedding of the sheet.
The conical singularity at $r=0$ breaks ``translational'' material symmetry of the sheet.  The remaining symmetries of the system are the ``rotational'' material symmetry of the sheet around the singularity, and the translational and rotational symmetries of space.  Thus, linear and angular momentum, and angular pseudomomentum, are conserved.  We will discuss and exploit both angular quantities in the following sections. 
Rather than explicitly rearranging the tangential projection of momentum balance into the pseudomomentum balance, we will obtain conservation laws directly from the boundary terms in the variation.

\subsection{Inertia: rotation and circumferential flow}\label{sec:conical_membranes}

An action for a perfectly flexible, inextensible sheet involves inertia and a constraint on the metric \cite{Guven13skirts},
\begin{align}
A = \int_{t_0}^{t_1}\!\!dt\int_{\mathcal{B}} \dArea\,\sqrt{\bar{a}/a}\,\left[\tfrac{1}{2}\rconfig{\rho}\dt{\bx}\cdot\dt{\bx} - \tfrac{1}{2}\sigma^{\alpha\beta}\left(\ga_{\alpha\beta} - \rconfig{\ga}_{\alpha\beta}\right)\right]\, , \label{conical_membrane_action}
\end{align}
where $\dArea = \sqrt{a}\,d\mc^1d\mc^2$. 
Performing a variation in position and material coordinates, and subsequently invoking $\sqrt{\bar{a}/a}\,\rconfig{\rho}=\rho$ and $\sqrt{\bar{a}/a}=1$, we obtain the variation in terms of the stress $\bff^\alpha$ and the (symmetric) material stress $T^\alpha_\gamma$,
\begin{align}
	&\delta A = \int_{t_0}^{t_1}\!\!dt\int_{\mathcal{B}}\dArea\, \bigg(\dt{\bigg[\rho\dt{\bx}\cdot\left(\delta\bx -\nabla_\gamma\bx\, \delta\mc^\gamma\right)\bigg]} + \nabla_\alpha\bigg(- \bff^\alpha\cdot\delta\bx +T^\alpha_\gamma \delta\mc^\gamma  \bigg) \nonumber \\
& \qquad\qquad\qquad\qquad \;\;\,+  \bigg[-\dt{\left(\rho\dt{\bx}\right)} + \nabla_\alpha\bff^\alpha\bigg]\cdot\left(\delta\bx - \nabla_\gamma\bx \,\delta\mc^\gamma\right)\bigg) \, , \label{inertial_membrane_full_variation} \\
&\bff^\alpha =\sigma^{\alpha\beta}\nabla_\beta\bx\, ,\quad
T^\alpha_\gamma = \bff^\alpha\cdot\nabla_\gamma\bx + \, \delta^\alpha_\gamma\left(\tfrac{1}{2}\rho\dt{\bx}\cdot\dt{\bx}\right) \, . \label{inertial_membrane_stress_material_stress}
\end{align}

\begin{figure}[h!]
	\includegraphics[width=7cm]{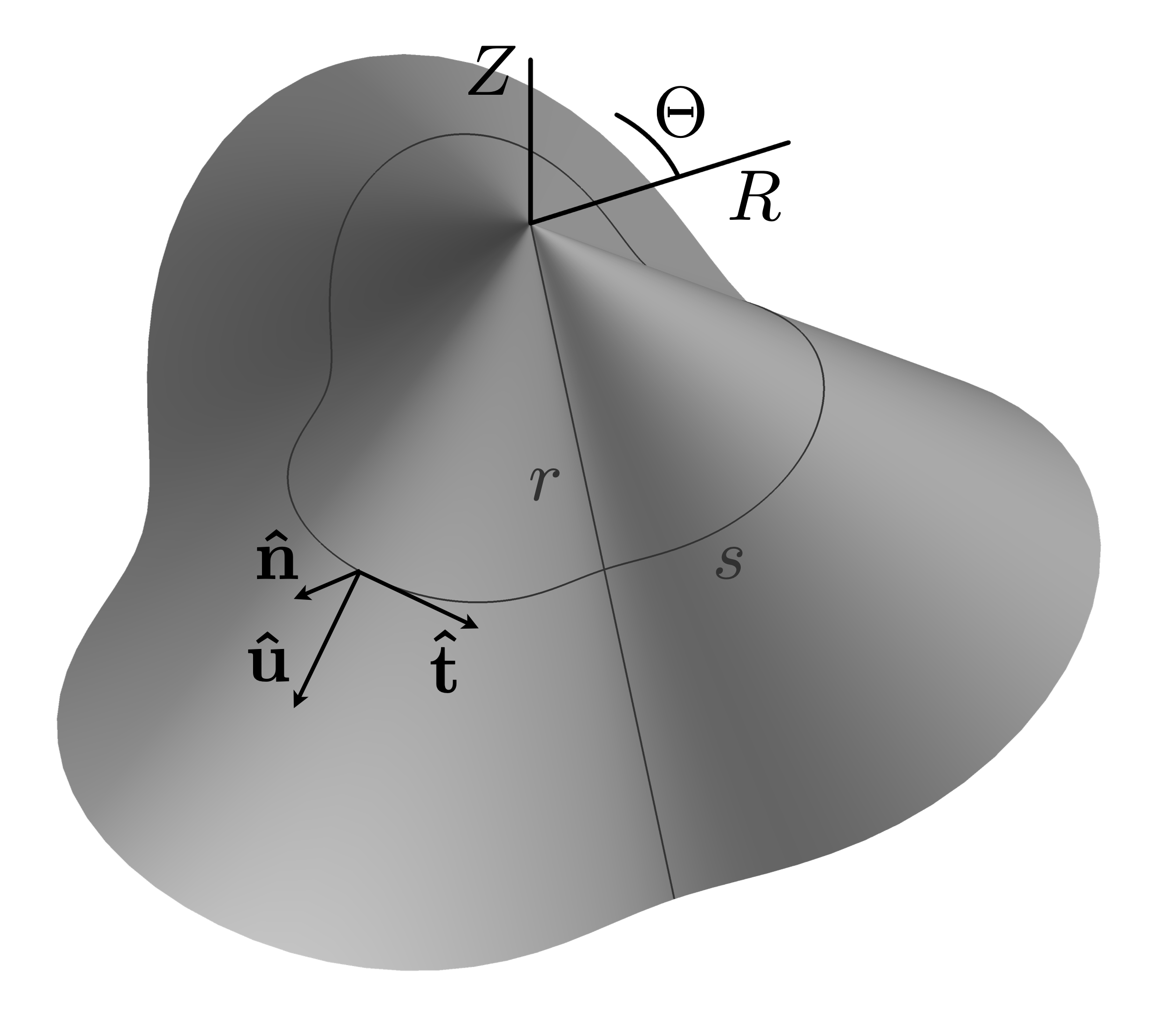}
	\captionsetup{margin=2cm}
	\vspace{-0.5cm}
	\caption{A conical surface with material coordinates $r, s$ and adapted frame $\uvc{u}, \uvc{t}, \uvc{n}$.  Material properties are symmetric in $s$.  Also shown are cylindrical coordinates $R, \Theta, Z$.  }
	\label{Fig:cone}
\end{figure}

We will be concerned with the general conservation laws associated with spatial and material symmetries $\delta\bx$ and $\delta\mc^\gamma$,
\begin{align}
	\dt{\left(\rho\dt{\bx}\cdot\delta\bx\right)} &= \nabla_\alpha\left(\bff^\alpha\cdot\delta\bx\right)\, ,\label{conical_membrane_physical_momentum_balance}\\
	\dt{\left(\rho\dt{\bx}\cdot\nabla_\gamma\bx\, \delta\mc^\gamma\right)}&=\nabla_\alpha\left( T^{\alpha}_{\gamma}\delta\mc^\gamma\right) \, .\label{conical_material_momentum_balance}
\end{align}
These we will apply to a prescribed equilibrium motion, consisting of rotation of a steady (generalized-)conical shape about a spatial axis $\uvc{Z}$ with an additional $s$-tangential flow superposed along the shape.
Defining spatial cylindrical coordinates using the axis of rotation, we can express the direction of the position vector as $\uvc{u} = R\uvc{R}+Z\uvc{Z}$.  The velocity is thus a combination of a rotation with angular velocity $\omega$ in the direction $\uvc{\Theta} = \uvc{Z}\times\uvc{R}$ and a tangential flow with ``angular velocity'' $\tau$ in the circumferential $s$ direction, $d_t\bx = \omega r R \uvc{\Theta} + \tau d_s\bx$, where $d_s\bx = r \uvc{t}$.  Figure \ref{Fig:cone} illustrates the two sets of coordinates and the adapted frame.

For such equilibria, there is a symmetry about the rotational axis, and the $Z$ component of the conserved angular momentum vector is itself conserved.  We start by inserting a small spatial rotational shift $\delta \bx = \epsilon \uvc{Z}\times\uvc{x} = \epsilon r R \uvc{\Theta}$ into \eqref{conical_membrane_physical_momentum_balance}.  
If, as in prior work \cite{Guven13skirts}, boundary conditions allow us to set off-diagonal terms $\sigma^{rs}=0$, we can simplify the right hand side to an $s$-derivative, because $\uvc{u}\cdot\uvc{\Theta}=0$ and the Christoffel symbol $\Gamma^s_{ss}=0$.
Additionally, because the configuration is steady, the quantities inside the outer time derivative on the left hand side change only through the tangential motion of material along the steady shape.  Thus, we may substitute $\tau d_s$ for the outer $d_t$.  Finally, as $\tau$ is uniform, it may be moved inside the $s$-derivative, and thus all terms can be grouped into a single conserved quantity,
\begin{align}
	0 = \epsilon r^2 d_s \left[ \uvc{t}\cdot\uvc{\Theta}\, R \left( \sigma^{ss}-\rho\tau^2 \right) - \rho\omega R^2\tau \right] \, . \label{conserved_angular_momentum}
\end{align}

A material ``rotation'' in two dimensions, or rather circumferential shift $\delta r = 0$, $\delta s = \epsilon$,  around the singularity $r=0$, will provide us with the only component of the angular material momentum.  Inserting this shift into \eqref{conical_material_momentum_balance} and following similar lines, noting that $T^r_s=0$ and $\sigma^s_s = r^2\sigma^{ss}$, we obtain
\begin{align}
	0 = \epsilon r^2 d_s \left[ \sigma^{ss} + \tfrac{1}{2}\rho\left( \omega^2 R^2 - \tau^2 \right) \right] \, . \label{conserved_angular_pseudomomentum}
\end{align}

The two conserved quantities bracketed in \eqref{conserved_angular_momentum} and \eqref{conserved_angular_pseudomomentum}, along with the flow parameter $\tau$, are sufficient to classify all such rotating, flowing equilibria of perfectly flexible conical sheets \cite{Guven13skirts}.  Eliminating the $\sigma^{ss}$ term, using the relation $R^2+Z^2=1$ (and its $s$-derivative), and noting that $\uvc{t}\cdot\uvc{R} = d_s R$, $\uvc{t}\cdot\uvc{Z} = d_s Z$, and $\left(\uvc{t}\cdot\uvc{R}\right)^2 + \left(\uvc{t}\cdot\uvc{\Theta}\right)^2 + \left(\uvc{t}\cdot\uvc{Z}\right)^2=1$, it is possible to construct an equation in terms of $R$ and three constants that is first order in $s$-derivatives of $R$.  We leave this as an exercise for the enthusiastic reader.
Alternatively, one could similarly construct a quadrature for Z, as in the prior work \cite{Guven13skirts}.  A related work on rotating, flowing strings \cite{HannaPendar16} used an alternate method in which the conserved $Z$-component of linear momentum was used to classify the equilibria.
This quantity can be captured in the present example by inserting a spatial translational shift $\delta \bx = \epsilon \uvc{Z}$ into \eqref{conical_membrane_physical_momentum_balance}.

\subsection{Bending elasticity}\label{sec:isometric_bending_surface}

Our next example is a static one in which inertia is neglected, but isometric flexure of the sheet is penalized with a bending energy quadratic in mean curvature \cite{guvenmuller2008} with associated uniform stiffness $\BH$, 
\begin{align}
A = \int_{\mathcal{B}} \dArea\,\sqrt{\bar{a}/a}\, \left(-\tfrac{1}{2}\right)\left[ 4 \BH H^2 + \sigma^{\alpha\beta}\left(\ga_{\alpha\beta} - \rconfig{\ga}_{\alpha\beta}\right)\right]\, ,\label{bending_surface_action}
\end{align}
where $2H = b^\alpha_\alpha$, and these components of the extrinsic curvature tensor may be defined either by $b_{\alpha\beta}=d_\beta \nabla_\alpha\bx\cdot\uvc{n}$ or through the Gauss-Weingarten relations $\nabla_\beta\nabla_\alpha\bx = b_{\alpha\beta}\uvc{n}$ and $\nabla_\alpha\uvc{n}=-b^\beta_\alpha\nabla_\beta\bx$.
The variation of the action may be written \cite{guvenmuller2008, Hanna19} in terms of the stress $\bff^\alpha$, the moment $\bm{\mu}$, and the material stress $T^\alpha_\gamma$, after invoking $\sqrt{\bar{a}/a} = 1$,
\begin{align}
&\delta A = \int_{\mathcal{B}}\dArea\, \bigg[\nabla_\alpha\left( - \bff^\alpha\cdot\delta\bx - \bm{\mu}\cdot\nabla^\alpha\delta\bx + T^{\alpha}_{\gamma}\delta\mc^\gamma \right) + \nabla_\alpha\bff^\alpha\cdot\left(\delta\bx - \nabla_\gamma\bx\,\delta\mc^\gamma\right)\bigg] \, , \label{bending_membrane_full_variation} \\
&\bff^\alpha = \sigma^{\alpha\beta}\nabla_\beta\bx+ 2 \BH\left( H \nabla^\alpha \uvc{n} - \nabla^\alpha H  \uvc{n} \right) \, , \nonumber \\
&\bm{\mu} = 2\BH H \uvc{n}\, , \quad
T^\alpha_\gamma = \bff^\alpha\cdot\nabla_\gamma\bx + \bm{\mu}\cdot\nabla^\alpha\nabla_\gamma\bx - 2 \BH H^2\delta^\alpha_\gamma \, .\label{bending_membrane_things}
\end{align}
Note that $\sigma^\alpha_\gamma =  \bff^\alpha\cdot\nabla_\gamma\bx + \bm{\mu}\cdot\nabla^\alpha\nabla_\gamma\bx$. 
For the conical parametrization $\bx=r\uvc{u}(s,t)$, the only nonzero curvature is $b_{ss} = -r k$, and thus
$H = - k/2r$.

For this system with bending and without inertia, the general conservation laws associated with spatial and material symmetries $\delta\bx$ and $\delta\mc^\gamma$ take the form
\begin{align}
\nabla_\alpha\left(\bff^\alpha \cdot\delta\bx + \bm{\mu}\cdot\nabla^\alpha\delta\bx \right) &=0\, ,\label{bending_surface_conservation_physical_momentum}\\
\nabla_\alpha\left( T^\alpha_\gamma\delta\mc^\gamma\right) &= 0\, .\label{bending_surface_conservation_material_momentum}
\end{align}

The rotational symmetry of the embedding space around any arbitrary constant axis $\uvc{D}$ gives rise to the conservation of angular momentum.
We insert a small shift of the form $\delta\bx = \epsilon \uvc{D}\times\bx$ in \eqref{bending_surface_conservation_physical_momentum} and express the result in terms of the conserved torque $\bfm^\alpha$ \cite{guvenmuller2008},
\begin{align}
\nabla_\alpha\left(\bx\times\bff^\alpha + \nabla^\alpha\bx\times\bm{\mu}\right) &\equiv \nabla_\alpha\bfm^\alpha = {\bf 0}\, .\label{bending_physical_angular_momentum}
\end{align}
The boundary quantity corresponding to the surface divergence $\nabla_\alpha$ is $\bfm^\alpha \nu_\alpha$, where $\nu_r = \pm 1$ and $\nu_s = \pm r$ are the components of the unit tangents normal $\pm \uvc{u}$ and $\pm \uvc{t}$.  However, if $\sigma^{rs}=0$ as before, the $\bfm^r$ term does not contribute to this quantity and we may, following \cite{guvenmuller2008}, consider the quantity $\bJ \equiv r^2\bfm^s$ which is conserved along the $s$ coordinate lines.

The material circumferential symmetry corresponding to $\delta r = 0$, $\delta s = \epsilon$ gives rise to the conservation of angular material momentum.  Using \eqref{bending_surface_conservation_material_momentum} as before, we find $d_s T^s_s = 0$ and, anticipating our result, define a constant $C$,
\begin{align}
	T^s_s = \sigma^s_s - 2\BH H^2 = -\BH\frac{C}{r^2} \, . \label{Cdefinition}
\end{align}

Making use of \eqref{Cdefinition}, we can express $\bJ$ as 
\begin{align}
\bJ&=  -\BH \left[ k\, \uvc{u} + \partial_s k\, \uvc{t} + \left(\tfrac{1}{2} k^2 + C\right)\,\uvc{n} \right] \, .\label{cone_boundary_spatial_angular_momentum}
\end{align}
Its squared modulus is a quadrature for $k$ that can be arranged into the same form as that of planar Euler \emph{elastica} \cite{guvenmuller2008}, thus classifying all static equilibria of inextensible conical sheets with quadratic mean curvature energy in terms of the ratio of magnitudes of the angular momentum and pseudomomentum \cite{SinghHanna19}.

\section{Other avenues}

Pseudomomentum is an under-utilized concept in many corners of continuum and structural mechanics.  Beyond the examples in this paper and in the recent monograph by O'Reilly \cite{oreilly2017}, we suggest that the most fruitful and interesting targets for the application of these approaches may be problems involving evolving or active matter \cite{OzakinYavari10}
or fluid-structure interactions, including general locomotion or the motion of triple points at capillary contact lines \cite{Masurel19} or during water entry of structures \cite{Pesce03}. 

\section{Conclusions}

We have presented a general variational framework, within which we have applied the balance of pseudomomentum to derive and interpret results on a variety of continua including elastic solids, ideal fluids, and thin structures.

\section*{Acknowledgments}
This work was supported by U.S. National Science Foundation grant CMMI-1462501. H.S. acknowledges partial support by Swiss National Science Foundation grant 200020$\_$182184 to J. H. Maddocks.
We thank A. Gupta and A. Yavari for helpful suggestions.

\appendix

\section{Referential description of simple elastic solid}\label{app:derivation_reference_coordinates}

Here we rederive some of the results of Section \ref{sec:first_gradient_theory} while working in the referential frame.  Defining $\F \equiv \nabla_i\bx\rconfig{\nabla}^i\rconfig{\bx}$ and $\F^{-1} \equiv \rconfig{\nabla}_i\rconfig\bx\nabla^i\bx$, the two sets of results can be translated into each other by use of the Piola transforms $J\nabla\cdot() = \rconfig\nabla\cdot\left(J\F^{-1}\cdot()\right)$ and $J^{-1}\rconfig\nabla\cdot() = \nabla\cdot\left(J^{-1}\F\cdot()\right)$, special cases of which are the Piola identities that hold when $()$ is unity, one of which was used in Section \ref{sec:physical_and_material_forces}.  

We begin by expressing the action using a reference volume integral, $A = \int_{t_0}^{t_1}\!dt \! \int_{\mathcal{B}}\dvolref \Lref\left({\rconfig{\bx},t;\bx,\dt{\bx},\F}\right)$.  Under transformations of the dependent and independent fields, the change in the action is 
 \begin{align}
 \delta A = \int_{t_0}^{t_1}\!\! dt \int_{\mathcal{B}} \dvolref \left[\dt{\left(\Lref\,\delta t\right)} + \rconfig{\nabla}_i\left(\Lref\,\delta \mc^i\right) +\shop{\delta}\Lref\right]\, ,\label{reference_action_shift}
 \end{align}
which may be arranged as
\begin{align}
\delta A = \int_{t_0}^{t_1}\!\! dt \int_{\mathcal{B}}\dvolref \left[\dt{\mathcal{Q}} + \rconfig{\nabla}\cdot\bm{\mathcal{J}}^{(R)} + \bm{\mathcal{E}}^{(R)}\cdot\shop{\delta}\bx\right]\, ,
\end{align}
with the Euler-Lagrange term
\begin{align}
\bm{\mathcal{E}}^{(R)}&= \frac{\partial\Lref}{\partial\bx} - \dt{\left(\frac{\partial\Lref}{\partial\dt{\bx}}\right)} - \rconfig{\nabla}\cdot\left(\left[\frac{\partial\Lref}{\partial\F}\right]^\mathrm{T}\right) \, ,
\end{align}
and charge and current terms
\begin{align}
\mathcal{Q} &= \frac{\partial\Lref}{\partial\dt{\bx}}\cdot\delta\bx + \left(\frac{\partial\Lref}{\partial\dt{\bx}}\cdot\F\right)\cdot\left(-\delta\rconfig{\bx}\right) + \left(\frac{\partial\Lref}{\partial\dt{\bx}}\cdot\dt{\bx} - \Lref\right) (-\delta t) \, ,\\
\bm{\mathcal{J}}^{(R)} &= \left[\frac{\partial\Lref}{\partial\F}\right]^\mathrm{T}\cdot\delta\bx + \left(\left[\frac{\partial\Lref}{\partial\F}\right]^\mathrm{T}\cdot\F - \Lref\, \bf{I}\right)\cdot\left(-\delta\rconfig{\bx}\right) + \left(\left[\frac{\partial\Lref}{\partial\F}\right]^\mathrm{T}\cdot\dt{\bx}\right)(-\delta t) \, .
\end{align}

As in Section \ref{sec:first_gradient_theory}, we obtain balance laws for momentum,
\begin{align}
\bm{\mathcal{E}}(\Lref) \equiv \frac{\partial\Lref}{\partial\bx} - \dt{\left(\frac{\partial\Lref}{\partial\dt{\bx}}\right)} - \rconfig{\nabla}\cdot\left(\left[\frac{\partial\Lref}{\partial\F}\right]^\mathrm{T}\right) &= \bm{0}\quad\text{on}\;\mathcal{B}\, ,\label{reference_physical_momentum}\\
\bm{\mathrm{\hat{\bar{n}}}}\cdot\left[\frac{\partial\Lref}{\partial\F}\right]^\mathrm{T}  &= \bm{0}\quad\text{on}\;\partial\mathcal{B}\\
\jump{-\bm{\mathrm{\hat{\bar{N}}}}\cdot\left[\frac{\partial\Lref}{\partial\F}\right]^\mathrm{T} + \bar U\frac{\partial\Lref}{\partial\dt{\bx}}}&=\bm{0}\quad\text{on}\;{\mathcal{S}}(t)
\end{align}
pseudomomentum,
\begin{align}
\bm{\mathcal{E}}(\Lref)\cdot\F \equiv \left[\frac{\partial\Lref}{\partial\bx} - \dt{\left(\frac{\partial\Lref}{\partial\dt{\bx}}\right)} - \rconfig{\nabla}\cdot\left(\left[\frac{\partial\Lref}{\partial\F}\right]^\mathrm{T}\right)\right]\cdot\F &= \bm{0}\quad\text{on}\;\mathcal{B}\, ,\label{reference_material_momentum}\\
\bm{\mathrm{\hat{\bar{n}}}}\cdot\left(\left[\frac{\partial\Lref}{\partial\F}\right]^\mathrm{T}\cdot\bF - \Lref\,\bf{I}\right) &= \bm{0}\quad\text{on}\;{\partial\mathcal{B}}\, ,\\
\jump{-\bm{\mathrm{\hat{\bar{N}}}}\cdot\left(\left[\frac{\partial\Lref}{\partial\F}\right]^\mathrm{T}\cdot\bF - \Lref\,\bf{I}\right) + \bar U\left(\frac{\partial\Lref}{\partial\dt{\bx}}\cdot\bF\right)}&=\bm{0}\quad\text{on}\;{\mathcal{S}}(t)\, ,
\end{align}
and energy,
\begin{align}
\bm{\mathcal{E}}(\Lref)\cdot\dt{\bx}\equiv \left[\frac{\partial\Lref}{\partial\bx} - \dt{\left(\frac{\partial\Lref}{\partial\dt{\bx}}\right)} - \rconfig{\nabla}\cdot\left(\left[\frac{\partial\Lref}{\partial\F}\right]^\mathrm{T}\right)\right]\cdot\dt{\bx} &= 0\quad\text{on}\;\mathcal{B}\, ,\label{reference_energy_balance}\\
\bm{\mathrm{\hat{\bar{n}}}}\cdot\left(\left[\frac{\partial\Lref}{\partial\F}\right]^\mathrm{T}\cdot\dt{\bx}\right)&=0\quad\text{on}\;{\partial\mathcal{B}}\, ,\\
\jump{-\bm{\mathrm{\hat{\bar{N}}}}\cdot\left(\left[\frac{\partial\Lref}{\partial\F}\right]^\mathrm{T}\cdot\dt{\bx}\right) + \bar U\left(\frac{\partial\Lref}{\partial\dt{\bx}} - \Lref\right)}&=0\quad\text{on}\;{\mathcal{S}}(t)\, ,
\end{align}
where $\bm{\mathrm{\hat{\bar{n}}}}$ and $\bm{\mathrm{\hat{\bar{N}}}}$ are unit normals to the external boundary and internal surface of discontinuity in the reference configuration.

The balance of momentum has the form
\begin{align}
\dt{\left(\frac{\partial\Lref}{\partial\dt{\bx}}\right)} + \rconfig{\nabla}\cdot\left(\left[\frac{\partial\Lref}{\partial\F}\right]^\mathrm{T}\right) &= \frac{\partial\Lref}{\partial\bx}  \, .
\end{align}
The balance of energy can also, with the help of \eqref{lagrangian_time_derivative}, be rearranged into
\begin{align}
\dt{\left(\frac{\partial\Lref}{\partial\dt{\bx}}\cdot\dt{\bx} - \Lref\right)} + \rconfig{\nabla}\cdot\left(\left[\frac{\partial\Lref}{\partial\F}\right]^\mathrm{T}\cdot\dt{\bx}\right) = -\frac{\partial\Lref}{\partial t}\, .
\end{align}
Finally, with the chain rule
\begin{align}
\rconfig{\nabla}\Lref = \frac{\partial\Lref}{\partial\rconfig{\bx}} + \frac{\partial\Lref}{\partial\bx}\cdot\F+ \frac{\partial\Lref}{\partial\dt{\bx}}\cdot\dt{\F} + \left[\frac{\partial\Lref}{\partial\F}\right]^\mathrm{T}:\rconfig{\nabla}\F\, ,\label{reference_material_chain_rule}
\end{align}
where the double contraction involves 
both legs (one referential and one present) of $\F$, 
we can rearrange the balance of pseudomomentum \eqref{reference_material_momentum} to obtain
\begin{align}
\dt{\left(\frac{\partial\Lref}{\partial\dt{\bx}}\cdot\F\right)} + \rconfig{\nabla}\cdot\left(\left[\frac{\partial\Lref}{\partial\F}\right]^\mathrm{T}\cdot\F - \Lref\, {\bf I}\right) = -\frac{\partial\Lref}{\partial\rconfig{\bx}}\, .\label{reference_material_momentum_balance_2}
\end{align}
The quantity appearing inside the referential divergence is the most commonly presented form of the Eshelby tensor \cite{eshelby75,maugin1993}.

\bibliographystyle{unsrt}

\begin{thebibliography}{10}

\bibitem{peierls1985}
R.~Peierls.
\newblock Momentum and pseudomomentum of light and sound.
\newblock In F.~Bassani, F.~Fumi, and M.~P. Tosi, editors, {\em Proceedings of
  the International School of Physics ``Enrico Fermi'': Highlights of
  Condensed-Matter Theory}, pages 237--255. North-Holland, Amsterdam, 1985.

\bibitem{rogula1966}
D.~Rogula.
\newblock Noether's theorem for a continuous medium interacting with external
  fields.
\newblock {\em Proceedings of Vibration Problems}, 7:337--344, 1966.

\bibitem{eshelby51}
J.~D. Eshelby.
\newblock The force on an elastic singularity.
\newblock {\em Phil. Trans. R. Soc. Lond. A}, 244:87--112, 1951.

\bibitem{eshelby1970}
J.~D. Eshelby.
\newblock Energy relations and the energy-momentum tensor in continuum
  mechanics.
\newblock In M.~F. Kanninen, W.~F. Adler, A.~R. Rosenfield, and R.~I. Jaffee,
  editors, {\em Inelastic behavior of solids}, pages 77--115. McGraw-Hill, New
  York, 1970.

\bibitem{eshelby75}
J.~D. Eshelby.
\newblock The elastic energy-momentum tensor.
\newblock {\em Journal of Elasticity}, 5:321--335, 1975.

\bibitem{landaulifshitzfields1971}
L.~D. Landau and E.~M. Lifshitz.
\newblock {\em The classical theory of fields}.
\newblock Pergamon Press, Oxford, 1971.

\bibitem{Ericksen95}
J.L. Ericksen.
\newblock Remarks concerning forces on line defects.
\newblock {\em Zeitschrift f{\"{u}}r angewandte Mathematik und Physik},
  46:S247--S271, 1995.

\bibitem{rice1968}
J.~R. Rice.
\newblock A path independent integral and the approximate analysis of strain
  concentration by notches and cracks.
\newblock {\em Journal of Applied Mechanics}, 35:379--386, 1968.

\bibitem{Cherepanov1967}
G.P. Cherepanov.
\newblock Crack propagation in continuous media.
\newblock {\em Journal of Applied Mathematics and Mechanics}, 31:503 -- 512,
  1967.

\bibitem{rogula77}
D.~Rogula.
\newblock Forces in material space.
\newblock {\em Archives of Mechanics}, 29:705--713, 1977.

\bibitem{herrmannalicia1981}
A.~Golebiewska Herrmann.
\newblock On conservation laws of continuum mechanics.
\newblock {\em International Journal of Solids and Structures}, 17:1--9, 1981.

\bibitem{herrmannalicia1982}
A.~Golebiewska Herrmann.
\newblock Material momentum tensor and path-independent integrals in fracture
  mechanics.
\newblock {\em International Journal of Solids and Structures}, 18:319--326,
  1982.

\bibitem{herrmannalicia1983}
A.~Golebiewska-Herrmann.
\newblock On the {L}agrangian formulation of continuum mechanics.
\newblock {\em Physica A}, 118:300--314, 1983.

\bibitem{kienzlerherrmann00}
R.~Kienzler and G.~Herrmann.
\newblock {\em Mechanics in Material Space}.
\newblock Springer, Berlin, 2000.

\bibitem{maugin1993}
G.~Maugin.
\newblock {\em Material Inhomogeneities in Elasticity}.
\newblock Chapman \& Hall, London, 1993.

\bibitem{maugin2011}
G.~A. Maugin.
\newblock {\em Configurational Forces}.
\newblock CRC Press, Boca Raton, 2011.

\bibitem{maugin95}
G.~A. Maugin.
\newblock Material forces: Concepts and applications.
\newblock {\em Applied Mechanics Reviews}, 48:213--245, 1991.

\bibitem{maugin2002}
G.~A. Maugin.
\newblock Recent advances in \emph{M}$^3$ (mechanics on the material manifold).
\newblock {\em Theoretical and Applied Mechanics}, 28-29:221--223, 2002.

\bibitem{oreilly2017}
O.~M. O'Reilly.
\newblock {\em Modeling Nonlinear Problems in the Mechanics of Strings and
  Rods}.
\newblock Springer, New York, 2017.

\bibitem{gurtin00}
M.~E. Gurtin.
\newblock {\em Configurational Forces as Basic Concepts of Continuum Physics}.
\newblock Springer, New York, 2000.

\bibitem{guidugli2001}
P.~Podio-Guidugli.
\newblock Configurational balance via variational arguments.
\newblock {\em Interfaces and Free Boundaries}, 3:223--232, 2001.

\bibitem{friedgurtin2005}
E.~Fried and M.~E. Gurtin.
\newblock The unifying nature of the configurational force balance.
\newblock In P.~Steinmann and G.~A. Maugin, editors, {\em Mechanics of Material
  forces}, pages 25--32. Springer, New York, 2005.

\bibitem{rajagopalsrinivasa2005}
K.~R. Rajagopal and A.~R. Srinivasa.
\newblock On the role of the {E}shelby energy-momentum tensor in materials with
  multiple natural configurations.
\newblock {\em Mathematics and Mechanics of Solids}, 10:3--24, 2005.

\bibitem{yavari06}
A.~Yavari, J.~E. Marsden, and M.~Ortiz.
\newblock On spatial and material covariant balance laws in elasticity.
\newblock {\em Journal of Mathematical Physics}, 47:042903, 2006.

\bibitem{edelen1981}
D.~G.~B. Edelen.
\newblock Aspects of variational arguments in the theory of elasticity: Fact
  and folklore.
\newblock {\em International Journal of Solids and Structures}, 17:729--740,
  1981.

\bibitem{MauginTrimarco92mixed}
G.~A. Maugin and C.~Trimarco.
\newblock Note on a mixed variational principle in finite elasticity.
\newblock {\em Rendiconti Lincei -- Matematica e Applicazioni}, Serie 9, Vol.
  3, n.1:69--74, 1992.

\bibitem{MauginTrimarco92fracture}
G.~A. Maugin and C.~Trimarco.
\newblock Pseudomomentum and material forces in nonlinear elasticity:
  variational formulations and application to brittle fracture.
\newblock {\em Acta Mechanica}, 94:1--28, 1992.

\bibitem{Sturrock62}
P.~A. Sturrock.
\newblock Energy and momentum in the theory of waves in plasmas.
\newblock In D.~Bershader, editor, {\em Plasma Hydromagnetics: Sixth Lockheed
  Symposium on Magnetohydrodynamics}, pages 47--57. Stanford University Press,
  Stanford, 1962.

\bibitem{gilbertmollow1968}
I.~H. Gilbert and B.~R. Mollow.
\newblock Momentum of longitudinal elastic vibrations.
\newblock {\em American Journal of Physics}, 9:822--825, 1968.

\bibitem{broer70}
L.~J.~F. Broer.
\newblock On the dynamics of strings.
\newblock {\em Journal of Engineering Mathematics}, 4:195--202, 1970.

\bibitem{knowlesSternberg1972}
J.~K. Knowles and E.~Sternberg.
\newblock On a class of conservation laws in linearized and finite elasticity.
\newblock {\em Archive for Rational Mechanics and Analysis}, 44:187, 1972.

\bibitem{fletcher1976}
D.~C. Fletcher.
\newblock Conservation laws in linear elastodynamics.
\newblock {\em Archive for Rational Mechanics and Analysis}, 60:329--353, 1976.

\bibitem{hill1986}
R.~Hill.
\newblock Energy-momentum tensors in elastostatics: Some reflections on the
  general theory.
\newblock {\em Journal of the Mechanics and Physics of Solids}, 34:305--317,
  1986.

\bibitem{nelson91}
D.~F. Nelson.
\newblock Momentum, pseudomomentum, and wave momentum: Toward resolving the
  {M}inkowski-{A}braham controversy.
\newblock {\em Physical Review A}, 44:3985--3996, 1991.

\bibitem{thellung1994}
A.~Thellung.
\newblock Momentum and quasimomentum in the physics of condensed matter.
\newblock In T.~Paszkiewicz and K.~Rapcewicz, editors, {\em Die Kunst of
  Phonons}, pages 15--32. Plenum Press, New York, 1994.

\bibitem{eckart1960}
C.~Eckart.
\newblock Variation principles of hydrodynamics.
\newblock {\em The Physics of Fluids}, 3:421--427, 1960.

\bibitem{newcomb1967}
W.~A. Newcomb.
\newblock Exchange invariance in fluid systems.
\newblock In {\em Proceedings of Symposia in Applied Mathematics Volume XVIII:
  Magneto-Fluid and Plasma Dynamics}, pages 152--161, 1967.

\bibitem{bretherton1970}
F.~P. Bretherton.
\newblock A note on {H}amilton's principle for perfect fluids.
\newblock {\em Journal of Fluid Mechanics}, 44:19--31, 1970.

\bibitem{salmon1988}
R.~Salmon.
\newblock Hamiltonian fluid mechanics.
\newblock {\em Annual Review of Fluid Mechanics}, 20:220--256, 1988.

\bibitem{Muller95}
P.~M{\" u}ller.
\newblock Ertel's potential vorticity theorem in physical oceanography.
\newblock {\em Reviews of Geophysics}, 33:67--97, 1995.

\bibitem{padhyemorrison1996}
N.~Padhye and P.~J. Morrison.
\newblock Fluid element relabeling symmetry.
\newblock {\em Physics letters A}, 219:287--292, 1996.

\bibitem{benjamin84}
T.~B. Benjamin.
\newblock Impulse, flow force and variational principles.
\newblock {\em IMA Journal of Applied Mathematics}, 32:3--68, 1984.

\bibitem{maddocksdichman1994}
J.~H. Maddocks and D.~J. Dichmann.
\newblock Conservation laws in the dynamics of rods.
\newblock {\em Journal of Elasticity}, 34:83--96, 1994.

\bibitem{healey1996}
T.~J. Healey.
\newblock Stability of axial motions of nonlinearly elastic loops.
\newblock {\em Zeitschrift f{\"{u}}r angewandte Mathematik und Physik},
  47:809--816, 1996.

\bibitem{mcIntyre1981}
M.~E. McIntyre.
\newblock On the `wave momentum' myth.
\newblock {\em Journal of Fluid Mechanics}, 106:331--347, 1981.

\bibitem{Shepherd90}
T.~G. Shepherd.
\newblock Symmetries, conservation laws, and {H}amiltonian structure in
  geophysical fluid dynamics.
\newblock {\em Advances in Geophysics}, 32:287--338, 1990.

\bibitem{BuehlerBOOK}
O.~B{\"{u}}hler.
\newblock {\em Waves and Mean Flows}.
\newblock Cambridge, New York, 2014.

\bibitem{cicconofridesimone2015}
G.~Cicconofri and A.~DeSimone.
\newblock A study of snake-like locomotion through the analysis of a flexible
  robot model.
\newblock {\em Proceedings of the Royal Society A}, 471:20150054, 2015.

\bibitem{dalcorso17}
F.~Dal Corso, D.~Misseroni, N.~M. Pugno, A.~B. Movchan, N.~V. Movchan, and
  D.~Bigoni.
\newblock Serpentine locomotion through elastic energy release.
\newblock {\em Journal of the Royal Society Interface}, 14:20170055, 2017.

\bibitem{Guven13skirts}
J.~Guven, J.~A. Hanna, and M.~M. M{\" u}ller.
\newblock Whirling skirts and rotating cones.
\newblock {\em New Journal of Physics}, 15:113055, 2013.

\bibitem{guvenmuller2008}
J.~Guven and M.~M. M{\" u}ller.
\newblock How paper folds: bending with local constraints.
\newblock {\em Journal of Physics A: Mathematical and Theoretical}, 41:055203,
  2008.

\bibitem{hill1951}
E.~L. Hill.
\newblock Hamilton's principle and the conservation theorems of mathematical
  physics.
\newblock {\em Reviews of Modern Physics}, 23:253--260, 1951.

\bibitem{rosen1972}
J.~Rosen.
\newblock Noether's theorem in classical field theory.
\newblock {\em Annals of Physics}, 69:349--363, 1972.

\bibitem{barbashov1983}
B.~M. Barbashov and V.~V. Nesterenko.
\newblock Continuous symmetries in field theory.
\newblock {\em Fortschritte der Physik}, 31:535--567, 1983.

\bibitem{LovelockRund88}
D.~Lovelock and H.~Rund.
\newblock {\em Tensors, Differential Forms, and Variational Principles}.
\newblock Dover, New York, 1988.

\bibitem{noethertavel1971}
E.~Noether.
\newblock Invariant variation problems.
\newblock {\em Transport Theory and Statistical Physics}, 1:183--207, 1971.
\newblock Translation by M. A. Tavel.

\bibitem{goldstein2001}
H.~Goldstein, C.~P. Poole, and J.~L. Safko.
\newblock {\em Classical Mechanics}.
\newblock Addison-Wesley, Boston, 2001.

\bibitem{gurtin:mechanics}
M.~E. Gurtin, E.~Fried, and L.~Anand.
\newblock {\em The Mechanics and Thermodynamics of Continua}.
\newblock Cambridge University Press, Cambridge, 2010.

\bibitem{eringen1980}
A.~C. Eringen.
\newblock {\em Mechanics of Continua}.
\newblock Robert E. Krieger Publishing Company, New York, 1980.

\bibitem{cermelliFried1997}
P.~Cermelli and E.~Fried.
\newblock The influence of inertia on the configurational forces in a
  deformable solid.
\newblock {\em Proceedings of the Royal Society of London A}, 453:1915--1927,
  1997.

\bibitem{markenscoff2006}
X.~Markenscoff.
\newblock Eshelby generalization for the dynamic \emph{{J}, {L}, {M}}
  integrals.
\newblock {\em Comptes Rendus M{\'{e}}canique}, 334:701--706, 2006.

\bibitem{nakamurashish1985}
T.~Nakamura, C.~F. Shih, and L.~B. Freund.
\newblock Computational methods based on an energy integral in dynamic
  fracture.
\newblock {\em International Journal of Fracture}, 27:1985, 229--243.

\bibitem{seligerwhitham1968}
R.~L. Seliger and G.~B. Whitham.
\newblock Variational principles in continuum mechanics.
\newblock {\em Proceedings of the Royal Society of London A}, 305:1--25, 1968.

\bibitem{kuzmin1983}
G.~A. Kuz'min.
\newblock Ideal incompressible hydrodynamics in terms of the vortex momentum
  density.
\newblock {\em Physics Letters}, 96A:88--90, 1983.

\bibitem{HannaPendar16}
J.~A. Hanna and H.~Pendar.
\newblock A conserved quantity in thin body dynamics.
\newblock {\em Physics Letters A}, 380:707--711, 2016.

\bibitem{thiffeault2001}
J.~Thiffeault.
\newblock Covariant time derivatives for dynamical systems.
\newblock {\em Journal of Physics A: Mathematical and General}, 34:5875--5885,
  2001.

\bibitem{caseynaghdi1991}
J.~Casey and P.~M. Naghdi.
\newblock On the {L}agrangian description of vorticity.
\newblock {\em Archive for Rational Mechanics and Analysis}, 115:1--14, 1991.

\bibitem{frischvillone2014}
U.~Frisch and B.~Villone.
\newblock Cauchy's almost forgotten {L}agrangian formulation of the {E}uler
  equation for 3d incompressible flow.
\newblock {\em The European Physical Journal H}, 39:325--351, 2014.

\bibitem{lamb1945}
H.~Lamb.
\newblock {\em Hydrodynamics}.
\newblock Dover, New York, 1945.

\bibitem{bennett2006}
A.~Bennett.
\newblock {\em Lagrangian Fluid Dynamics}.
\newblock Cambridge University Press, Cambridge, 2006.

\bibitem{cherepanov1977}
G.~P. Cherepanov.
\newblock Invariant {$\Gamma$}-integrals and some of their applications in
  mechanics.
\newblock {\em Journal of Applied Mathematics and Mechanics}, 41:399--412,
  1977.

\bibitem{atilgan1997}
A.~R. Atilgan.
\newblock Analogy between dislocation mechanics and aerodynamics.
\newblock {\em Journal of Applied Mathematics and Mechanics}, 77:631--633,
  1997.

\bibitem{moffatt1969}
H.~K. Moffatt.
\newblock The degree of knottedness of tangled vortex lines.
\newblock {\em Journal of Fluid Mechanics}, 35:117--129, 1969.

\bibitem{yahalom1995}
A.~Yahalom.
\newblock Helicity conservation via the {N}oether theorem.
\newblock {\em Journal of Mathematical Physics}, 36:1324--1327, 1995.

\bibitem{kienzlerherrmann1986}
R.~Kienzler and G.~Herrmann.
\newblock On material forces in elementary beam theory.
\newblock {\em Journal of Applied Mechanics}, 53:561--564, 1986.

\bibitem{bigoni15}
D.~Bigoni, F.~Dal~Corso, F.~Bosi, and D.~Misseroni.
\newblock Eshelby-like forces acting on elastic structures: {T}heoretical and
  experimental proof.
\newblock {\em Mechanics of Materials}, 80:368--374, 2015.

\bibitem{bigoni14}
D.~Bigoni, F.~Dal~Corso, D.~Misseroni, and F.~Bosi.
\newblock Torsional locomotion.
\newblock {\em Proceedings of the Royal Society A}, 470:20140599, 2014.

\bibitem{oreilly07}
O.~M. O'Reilly.
\newblock A material momentum balance law for rods.
\newblock {\em Journal of Elasticity}, 86:155--172, 2007.

\bibitem{oreilly2015eshelby}
O.~M. O'Reilly.
\newblock Some perspectives on {E}shelby-like forces in the elastica arm scale.
\newblock {\em Proceedings of the Royal Society A}, 471:20140785, 2015.

\bibitem{Hanna18}
J.~A. Hanna, H.~Singh, and E.~G. Virga.
\newblock Partial constraint singularities in elastic rods.
\newblock {\em Journal of Elasticity}, 133:105--118, 2018.

\bibitem{SinghHanna19}
H.~Singh and J.~A. Hanna.
\newblock On the planar \emph{elastica}, stress, and material stress.
\newblock {\em Journal of Elasticity}, 136(1):87--101, 2019.

\bibitem{Hanna19}
J.~A. Hanna.
\newblock Some observations on variational elasticity and its application to
  plates and membranes.
\newblock {\em Zeitschrift f{\"{u}}r angewandte Mathematik und Physik}, 70:76,
  2019.

\bibitem{OzakinYavari10}
A.~Ozakin and A.~Yavari.
\newblock A geometric theory of thermal stresses.
\newblock {\em Journal of Mathematical Physics}, 51:032902, 2010.

\bibitem{Masurel19}
R.~Masurel, M.~Roch{\'{e}}, L.~Limat, I.~Ionescu, and J.~Dervaux.
\newblock Elastocapillary ridge as a noninteger disclination.
\newblock {\em Physical Review Letters}, 122:248004, 2019.

\bibitem{Pesce03}
C.~P. Pesce.
\newblock The application of {L}agrange equations to mechanical systems with
  mass explicitly dependent on position.
\newblock {\em Journal of Applied Mechanics}, 70:751--756, 2003.

\end{thebibliography}

\end{document}